\documentclass{blois}
\pdfoutput=1
\title{Effect of deep gain layer and Carbon infusion on LGAD radiation hardness}

\usepackage{HGTD_TDR-defs}
\usepackage{multirow}
\usepackage{rotating}
\usepackage{mathrsfs}
\usepackage{booktabs}
\usepackage{longtable}
\usepackage{placeins}
\usepackage{amssymb}
\usepackage{upgreek}
\usepackage{svg}
\usepackage{comment}
\usepackage{pdfpages}
\usepackage{siunitx}
\usepackage{lineno}

\usepackage{cleveref}
\crefname{equation}{Eq.}{Eqs.}
\crefname{figure}{Fig.}{Figs.}
\crefname{table}{Tab.}{Tabs.}
\crefname{section}{Sec.}{Secs.}

\newcommand*{\ATLASLATEXPATH}{latex/}
\usepackage{\ATLASLATEXPATH atlasphysics}

\begin{document}
\maketitle
{\small \centering
            R~Padilla, C.~Labitan, Z.~Galloway, C.~Gee, S.~M.~Mazza\footnote{Corresponding author, simazza@ucsc.edu}, \\ 
            F.~McKinney-Martinez, H.~F.-W.~Sadrozinski, A.~Seiden, B.~Schumm,\\
            M.~Wilder, Y.~Zhao, H.~Ren, Y.~Jin, M.~Lockerby,\\
            V.~Cindro, G.~Kramberger, I.~Mandiz, M.~Mikuz, M.~Zavrtanik,\\
            R.~Arcidiacono, N.~Cartiglia, M.~Ferrero, M.~Mandurrino, V.~Sola, A.~Staiano\\
}

\begin{abstract}
The properties of \SI{50}{\micro\metre} thick Low Gain Avalanche Diode (LGAD) detectors manufactured by Hamamatsu photonics (HPK) and Fondazione Bruno Kessler (FBK) were tested before and after irradiation with neutrons.
Their performance were measured in charge collection studies using $\beta$-particles from a 90Sr source and in capacitance-voltage scans (C-V) to determine the bias to deplete the gain layer.
Carbon infusion to the gain layer of the sensors was tested by FBK in the UFSD3 production.
HPK instead produced LGADs with a very thin, highly doped and deep multiplication layer. 
The sensors were exposed to a 1-MeV neutron equivalent ($\mathrm{n}_\mathrm{eq}/\mathrm{cm}^2$) neutron fluence from \fluence{4}{14} to \fluence{4}{15}. 
The collected charge and the timing resolution were measured as a function of bias voltage at -30C, furthermore the profile of the capacitance over voltage of the sensors was measured. 
The deep gain layer (HPK-3.2) and the Carbon infusion (FBK3+C) both show reasonable performance up to \fluence{3}{15}. Furthermore the correlation between the time resolution and the gain is found to be independent from the sensor type, given the same thickness, and from the fluence. Finally a correlation is shown between the bias voltage to deplete the gain layer and the bias voltage needed to reach a certain amount of gain in the sensor.
\end{abstract}

\section{Introduction}

Low Gain Avalanche Detectors (LGADs) are thin (20 to 60 $\mu m$) n-on-p silicon sensors with modest internal gain (typically 5 to 50) and exceptional time resolution (17~ps to 50~ps)~\cite{bib:LGAD,bib:MarTorino,bib:UFSD300umTB}. 
LGADs were first developed by the Centro Nacional de Microelectrónica (CNM) Barcelona, in part as a RD50 Common Project \cite{rd50}. The internal gain is due to a highly doped p+ region (called multiplication or gain layer) just below the n-type implants of the electrodes.
The multiplication layer is up to a few microns thick, while the rest of the active area is referred to as the bulk.
In the next section the multiplication layer will be referred as ``thin'' if its thickness perpendicular to the junction is less than in a standard LGAD, ``highly doped'' if the maximum doping concentration in the gain layer is higher in respect to a standard LGAD and ``deep'' if the distance of the multiplication layer from the top surface is larger than in a standard LGAD.
Standard LGADs are defined as LGADs from previous prototype productions.
Thanks to their extraordinary properties LGADs establish a new paradigm for space-time particle tracking~\cite{bib:4Dtracking}. 

The first application of LGADs are planned for the High Luminosity LHC (HL-LHC~\cite{bib:hllhc}), where the extreme pileup conditions will lower the efficiency for tracking and vertexing of the inner tracking detector in the region close to the beam-pipe. Therefore, to maintain the performance, LGAD based timing layers are foreseen in the forward region of both the ATLAS and the CMS experiments. 
The two projects are called respectively the High Granularity Timing Detector (HGTD)~\cite{Collaboration:2623663} and the End-cap Timing Layer (ETL)~\cite{CMS:2667167}.
At HL-LHC, LGADs would be of moderate segmentation (1.3~mm~x~1.3~mm) and will have to face challenging radiation requirements, with fluences up to few \fluence{1}{15} and doses up to few MGy.

LGADs from several vendors have been tested extensively during the last few years. LGAD sensors have been proven to be able to reach a time resolution, before radiation damage, between 17~ps and 50~ps depending on thickness and doping profile. 
These measurements are in agreement with the simulation program Weightfield2 (WF2) \cite{WF2}. 
Previous studies on LGAD sensors from different vendors are reported in \cite{Mazza:2018jiz,Ferrero:2018fen,bib:HPKirradiation35vs50,bib:HPKirradiationGalloway,bib:HGTDBeamTest}. 
In all cited cases, both the timing resolution and the gain deteriorate with radiation damage due to the acceptor removal mechanism~\cite{Terada:1996hf,bib:LGADradiationGregor}, which reduces the effective doping concentration in the gain layer.
In the following sections it will be shown that the performance loss from radiation damage can be partly recovered by increasing the bias voltage applied to the sensor and that optimized sensor design can increase the recovery of performance after irradiation.

\section{Sensor types and electrical properties}

The four sensor types under study include two types by HPK, from a shared ATLAS-CMS LGAD prototype production, and two types by FBK from the INFN-funded production run called UFSD3. \cref{tab:sensortypes} shows the basic parameters of the four sensor types.
The geometry of the sensors tested is either single pads (HPK only) or 2x2 arrays (HPK and FBK). 
The pad dimension in all types is \SI{1.3}{\milli\metre}x\SI{1.3}{\milli\metre}.

In a deeper gain layer, such as that of HPK-3.2, the recovery of the gain due to an increase of the bias voltage is more pronounced since it is acting on a longer distance.
This design, therefore, is not more radiation resistance because the acceptor removal mechanism is slower, but because the bias voltage has stronger recovery capability. 

The two FBK sensor types are identical with the exception of FBK3+C having carbon added to the gain layer. 
Carbon was proven to increase the radiation hardness of LGADs because part of the Si interstitials responsible for Boron removal in the multiplication layer are de-activated by carbon capture.
However, since carbon is electrically inactive, its addition does not affect the sensor performance before irradiation.
This effect was already observed in the past FBK production FBK-UFSD2~\cite{Mazza:2018jiz,Ferrero:2018fen}.

\begin{table}[tb]
\centering
\begin{tabular}{|c|c|c|c|c|c|c|}
\hline
Manufacturer & Type & Active  & Physical  &  $V_{FD}$ &  $V_{BD}$ & Carbon\\
 &  &  Thickness &  Thickness &   & & \\
\hline
HPK & HPK-3.1 & \SI{50}{\micro\metre} & \SI{300}{\micro\metre} & 50~V & 250~V & no\\
HPK & HPK-3.2 & \SI{50}{\micro\metre} & \SI{300}{\micro\metre} & 70~V & 120~V & no\\
FBK & FBK3+C & \SI{55}{\micro\metre} & \SI{500}{\micro\metre} & 25~V & 400~V & yes\\
FBK & FBK3noC & \SI{55}{\micro\metre} & \SI{500}{\micro\metre} & 25~V & 400~V & no\\
\hline
\end{tabular}
\caption{Parameters of the detectors under study (at +20C), including Full Depletion Voltage ($V_{FD}$) and Breakdown Voltage ($V_{BD}$)}
\label{tab:sensortypes}
\end{table}

\begin{figure}[htbp]
\centering
\includegraphics[width=0.49\textwidth]{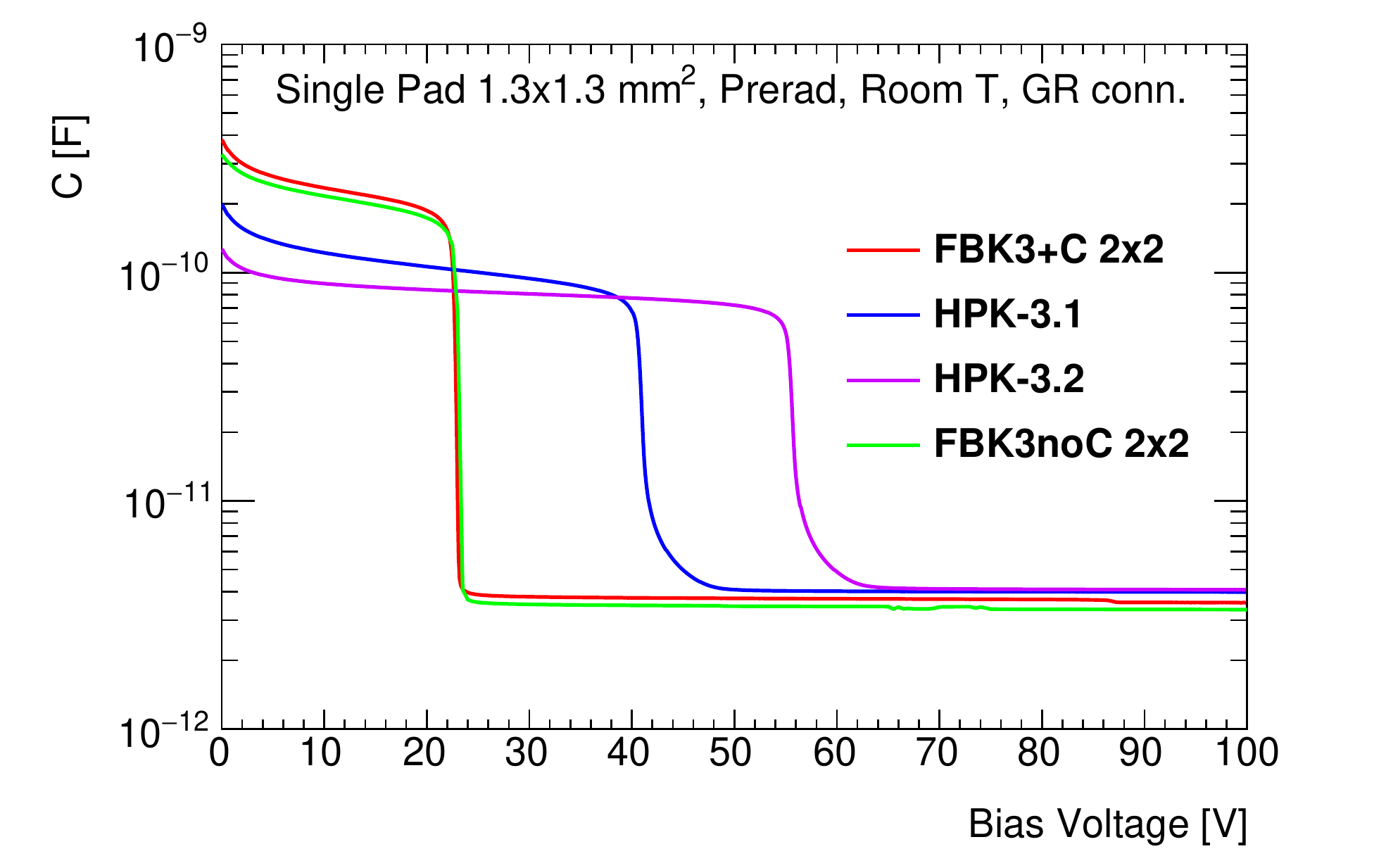}
\includegraphics[width=0.49\textwidth]{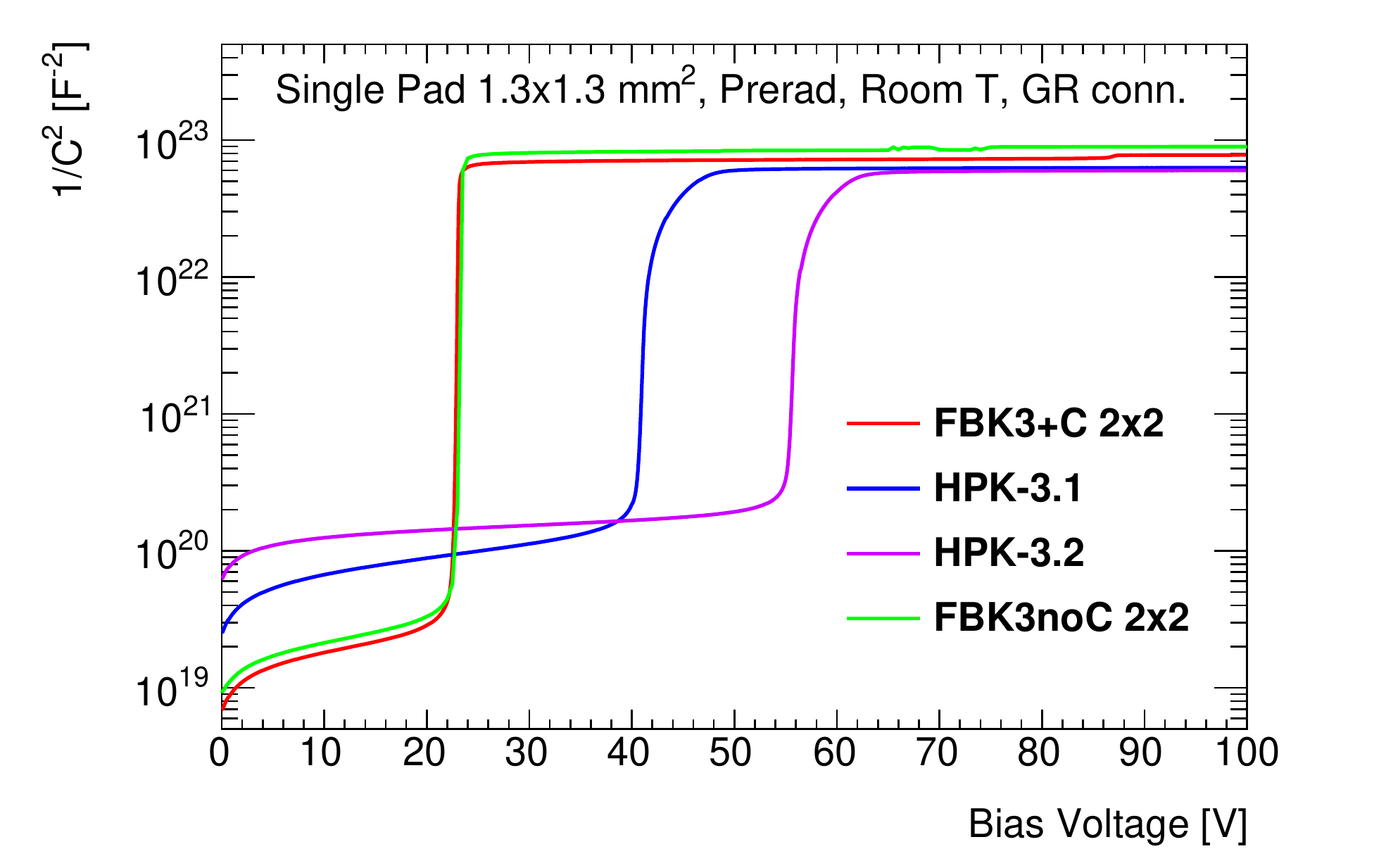}
\caption{C-V scan (Left) and $1/C^2$ distribution (right) of HPK and FBK3 sensors before irradiation.}
\label{fig:CV_doping}
\end{figure}

Basic electrical measurements on the sensors were done using a probe station equipped with needle contacts. The current-voltage (I-V) and the capacitance-voltage (C-V) scans were performed for all detectors.
The I-V measurements are used to evaluate the Breakdown Voltage ($V_{BD}$) in \cref{tab:sensortypes}, while the C-V are used to evaluate the Full Depletion Voltage ($V_{FD}$).
The C-V measurements were taken at \SI{10}{\kilo\hertz} for unirradiated sensors and at \SI{1}{\kilo\hertz} for irradiated sensors.
The measured C-V curves for all sensor types before irradiation are shown in \cref{fig:CV_doping} (left); the curves exhibit a strong fall-off at the bias where the gain layer depletes after which the C-V curve quickly reaches the saturation value revealing a lightly doped bulk.
The initial capacitance at low voltage (e.g. the capacitance value around 10~V for all sensor types) is an indication of the depth of the gain layer; since the active area is the same a lower initial capacitance is an indication of a deeper gain layer.

Studying the $1/C^2$ curve, \cref{fig:CV_doping} (right), three sections of the curve are common to all sensor types: a first flat section which corresponds to the depletion of the gain layer, then a fast rising section which corresponds to the depletion of the bulk and finally a flat constant part corresponding to the fully depleted sensor.
The intercept of the extensions of the fast rising section of the curve and the first flat section of the curve is called the ``foot'' ($V_{GL}$) and is proportional to the doping concentration in the multiplication layer multiplied by its depth.
The slope of the fast rising part of the curve is proportional to the doping concentration of the bulk.

Additional information can be extracted from the $1/C^2$ curve, knowing that the doping concentration of the gain layer of all types is similar.
It can be seen that the gain layer of FBK3+C and FBK3noC sensors have the same profile and have the lowest $V_{GL}$ (around 25~V) which correspond to a shallow gain layer.
Then HPK-3.1 has a higher $V_{GL}$ (around 40~V) corresponding to a deeper gain layer, finally HPK-3.2 has the highest $V_{GL}$ (around 55~V) corresponding to the deepest layer.

\section{Neutron irradiation at JSI}

The LGADs were irradiated without bias in the JSI research reactor of TRIGA type in Ljubljana, which has been used successfully in the past decades to support the development of radiation hard sensors~\cite{JSI_facility} with fluences between \fluence{4}{14} to \fluence{4}{15}. 
The neutron energy spectrum and flux are well known.
The fluence is quoted in 1 MeV equivalent neutrons per cm$^2$ ($n_{eq}/cm^2$) by using Non Ionizing Energy Loss (NIEL) scaling.


After irradiation, the devices were annealed for 80 min at 60~C. Afterward the devices were kept in cold storage at -20~C to reduce further annealing.

\section{Beta-scope setup and data analysis}

The charge collection experimental setup at SCIPP (Santa Cruz Institute for Particle Physics, University of California Santa Cruz) relies on a 90Sr beta-source, with a setup described in detail in \cite{Mazza:2018jiz,bib:HPKirradiation35vs50,bib:HPKirradiationGalloway}.
The tested LGAD, defined as device under test (DUT), is mounted on a fast analog electronic board (up to 2~GHz bandwidth) digitized by a GHz bandwidth digital scope.
The trigger, which acts as a time reference, is also mounted on a fast electronic board, and it is provided by a second HPK LGAD with time resolution of 17ps.
The electronic boards are mounted on a frame that aligns DUT and trigger to the 90Sr source.
The system is housed in a climate chamber allowing operations of irradiated sensors at temperatures down to -30C in a dry environment. Sensors before irradiation were tested both at 20C and -30C.

The digital oscilloscope records the full waveform of both trigger and DUT in each event, so the complete event information is available for offline analysis.
The data analysis follows the steps listed in \cite{Mazza:2018jiz,bib:HPKirradiation35vs50,bib:HPKirradiationGalloway} and proceeds as follows. 
The first step is a selection: for a valid trigger pulse, the signal amplitude Pmax of the DUT should not be saturated by either the scope or the read-out chain. 
To eliminate the contributions from non-gain events or noise, the time of the pulse maximum has to fall into a window of 1 ns centered on the trigger pulse.
The selected event waveforms are then analyzed to calculate the distribution of the pulse maximum, the collected charge, the gain, the rise and fall time and the time resolution.

For the calculation of the time resolution a Constant Fraction Discriminator (CFD) of 50\% is used to evaluate the time of arrival for the DUT and a CFD of 20\% for the time of arrival of the trigger. 
Then the distribution of the time differences between DUT and trigger is built
the $\sigma$ of a gaussian fit to the distribution is taken as the time resolution.
The time resolution of the DUT is then calculated by removing in quadrature the time resolution of the trigger, which is known\footnote{The time resolution for the trigger LGAD was measured by pairing two identical LGADs}.

The area of the pulse is evaluated for the DUT, subtracting the subsequent undershoot, then it is divided by the trans-impedence of the amplifier system (\SI{4700}{\ohm}) to calculate the collected charge.
The gain of the DUT LGAD is calculated dividing the collected charge by the collected charge of a same thickness PiN diode with no gain. 
The PiN diode charge is calculated with the Weightfield2~\cite{WF2} simulation software tuned with measurements of irradiated PiN diodes as explained with more detail in~\cite{Mazza:2018jiz}.

\section{Performance of unirradiated and irradiated sensors}

\begin{figure}[htbp]
\centering
\includegraphics[width=0.49\textwidth]{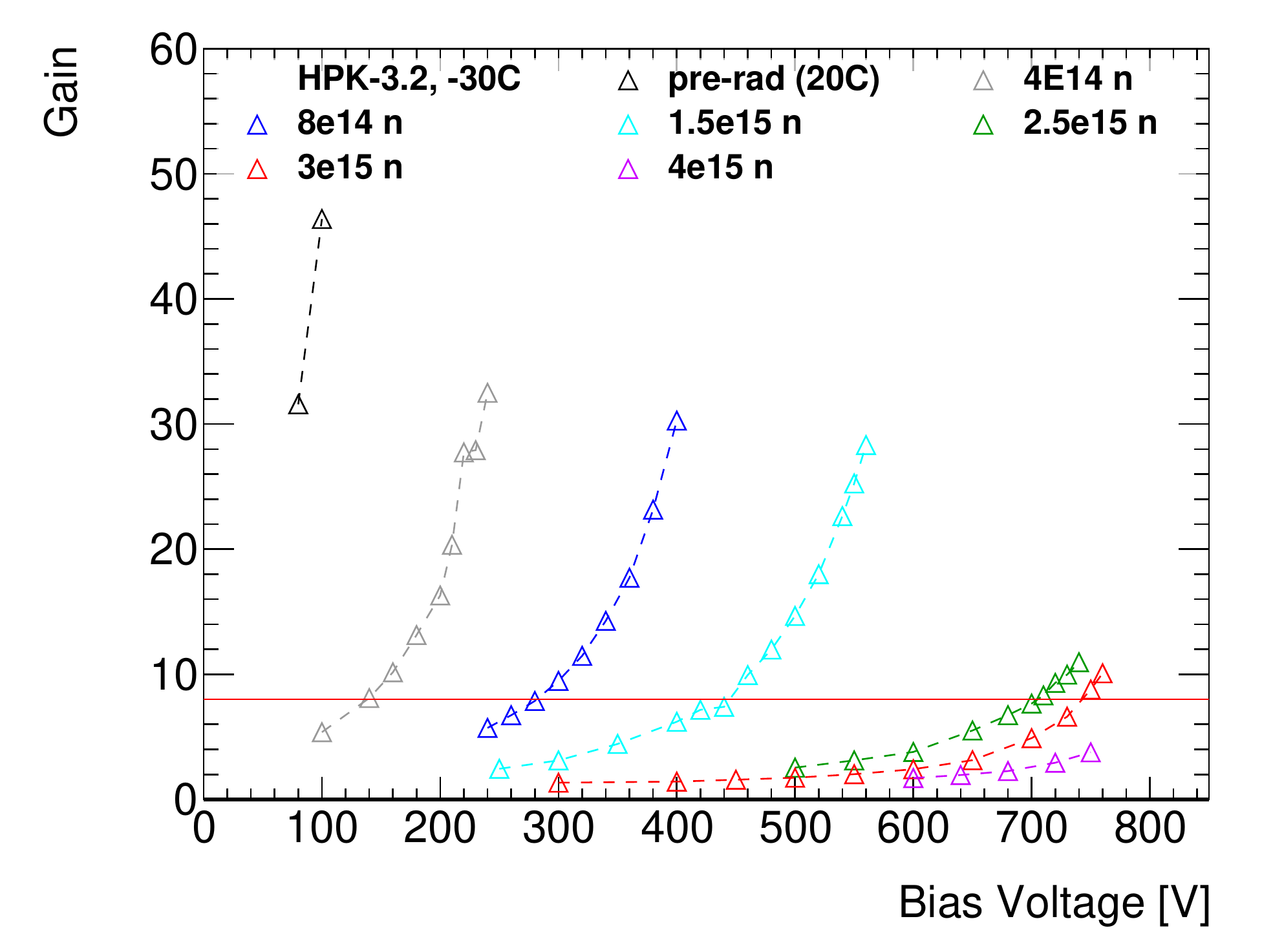}
\includegraphics[width=0.49\textwidth]{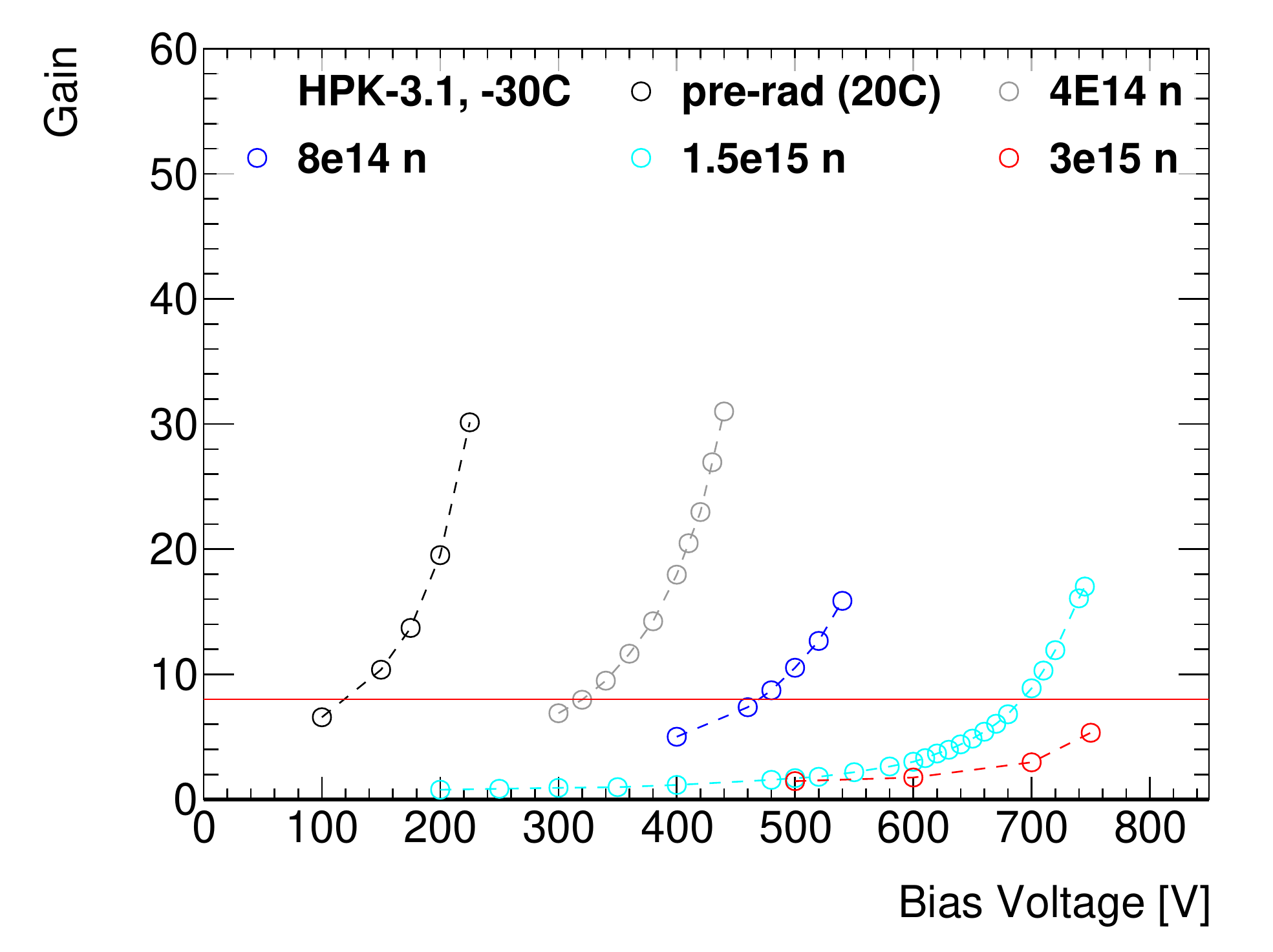}\\
\includegraphics[width=0.49\textwidth]{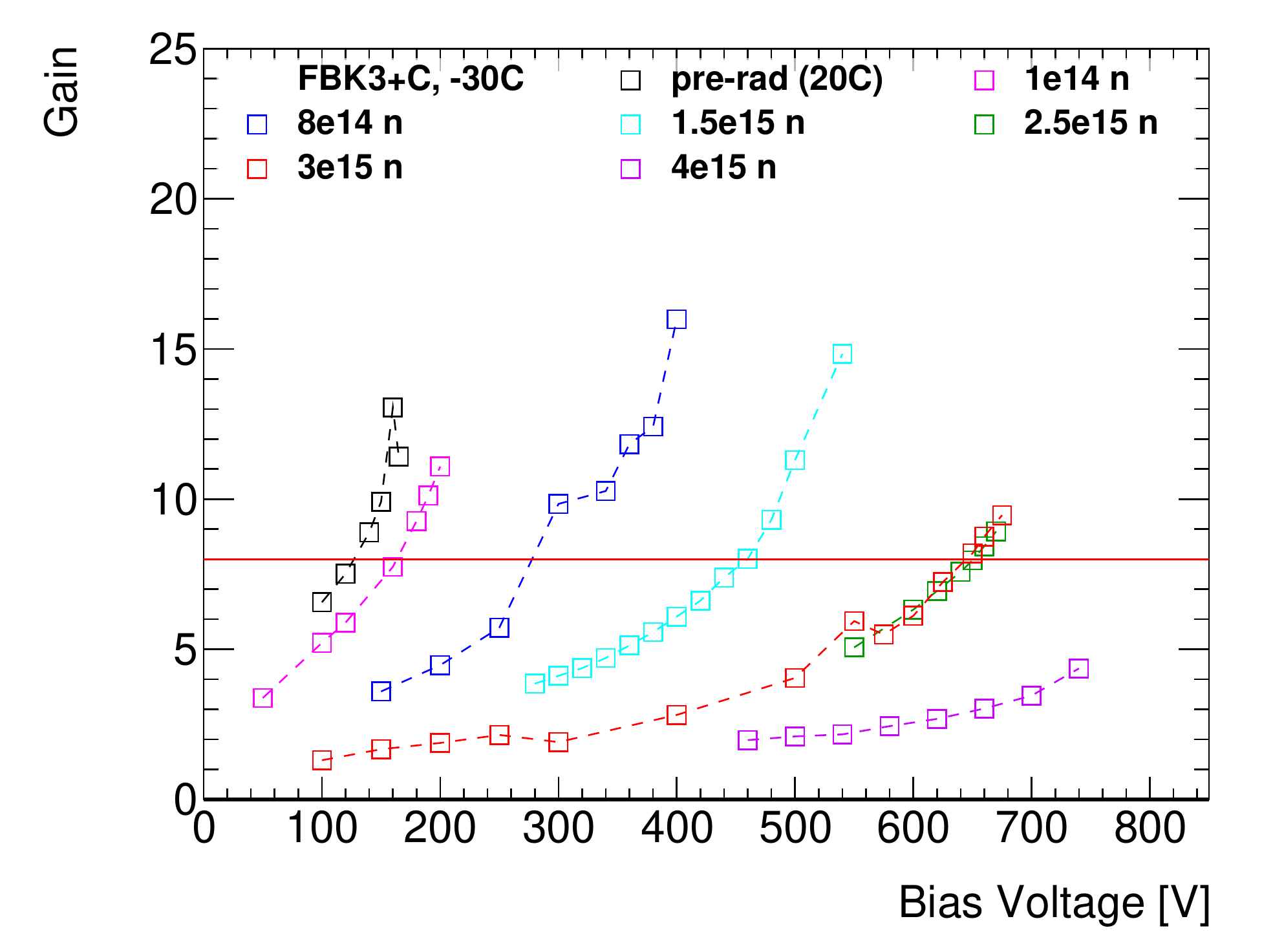}
\includegraphics[width=0.49\textwidth]{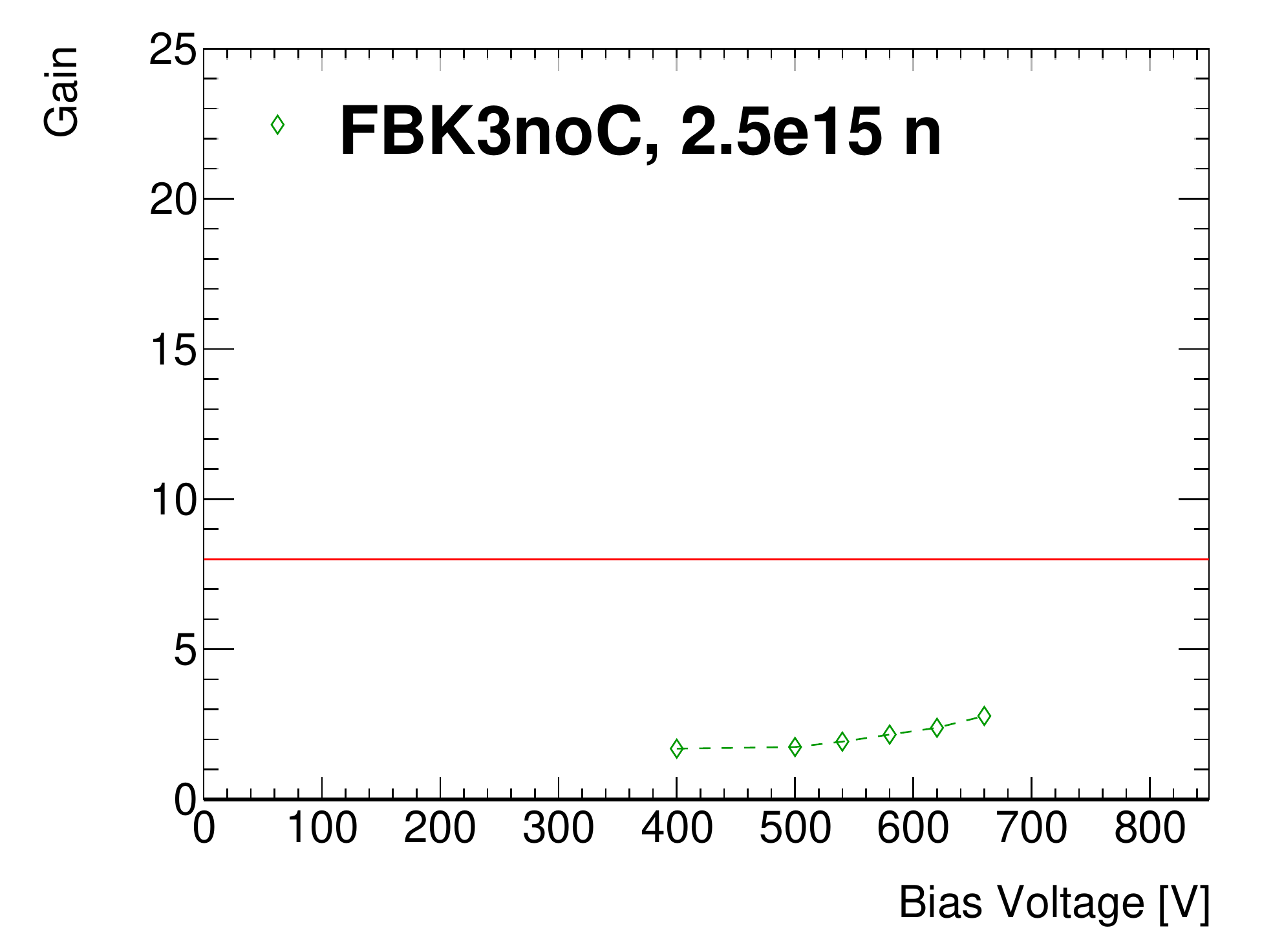}
\caption{Gain as a function of bias voltage for HPK-3.2 (top-left), HPK-3.1 (top-right), FBK3+C (bottom-left) and FBK3noC (bottom-right) sensors. The red horizontal line is for comparison purposes and represents a gain of 8. The plots for HPK-3.2/HPK-3.1 have a different vertical scale than for FBK3+C/FBK3noC.}
\label{fig:CC_beta}
\end{figure}

Sensors were tested with the Sr90 setup described in the previous section before and after irradiation with neutrons at JSI. 
The relevant measured parameters are the gain,
which is proportional to the collected charge, and the time resolution.
As seen in \cref{fig:CC_beta} for HPK-3.2, HPK-3.1, FBK3+C and FBK3noC the gain decreases with the fluence but can be partially restored by increasing the bias voltage applied.
However it was not possible to recover in this way a substantial gain at fluences above \fluence{3}{15}.

The performance of these sensors can be summarized as follows.
HPK-3.2 shows a very high gain before irradiation and up to \fluence{3}{15} it can still reach a gain of 8.
HPK-3.1 shows a similar performance (although at higher voltages) before irradiation but can reach a gain of 8 only up to \fluence{1.5}{15}.
FBK3+C has a lower overall gain but it can maintain it up to a fluence of \fluence{3}{15}. 
FBK3noC data is shown only after \fluence{2.5}{15} of irradiation because of measurement availability, at this fluence the sensor does not show gain. At the same fluence of \fluence{2.5}{15} the FBK sensor with Carbon (FBK3+C) still maintains a reasonable gain.
Finally, no sensor shows gain over 4 after a fluence of \fluence{4}{15}.
In \cref{sec:direct_comp}, \cref{fig:CC_compare_beta} the direct comparison of HPK-3.1/HPK-3.2 and FBK3+C/FBK3noC is shown.

\begin{figure}[htbp]
\centering
\includegraphics[width=0.49\textwidth]{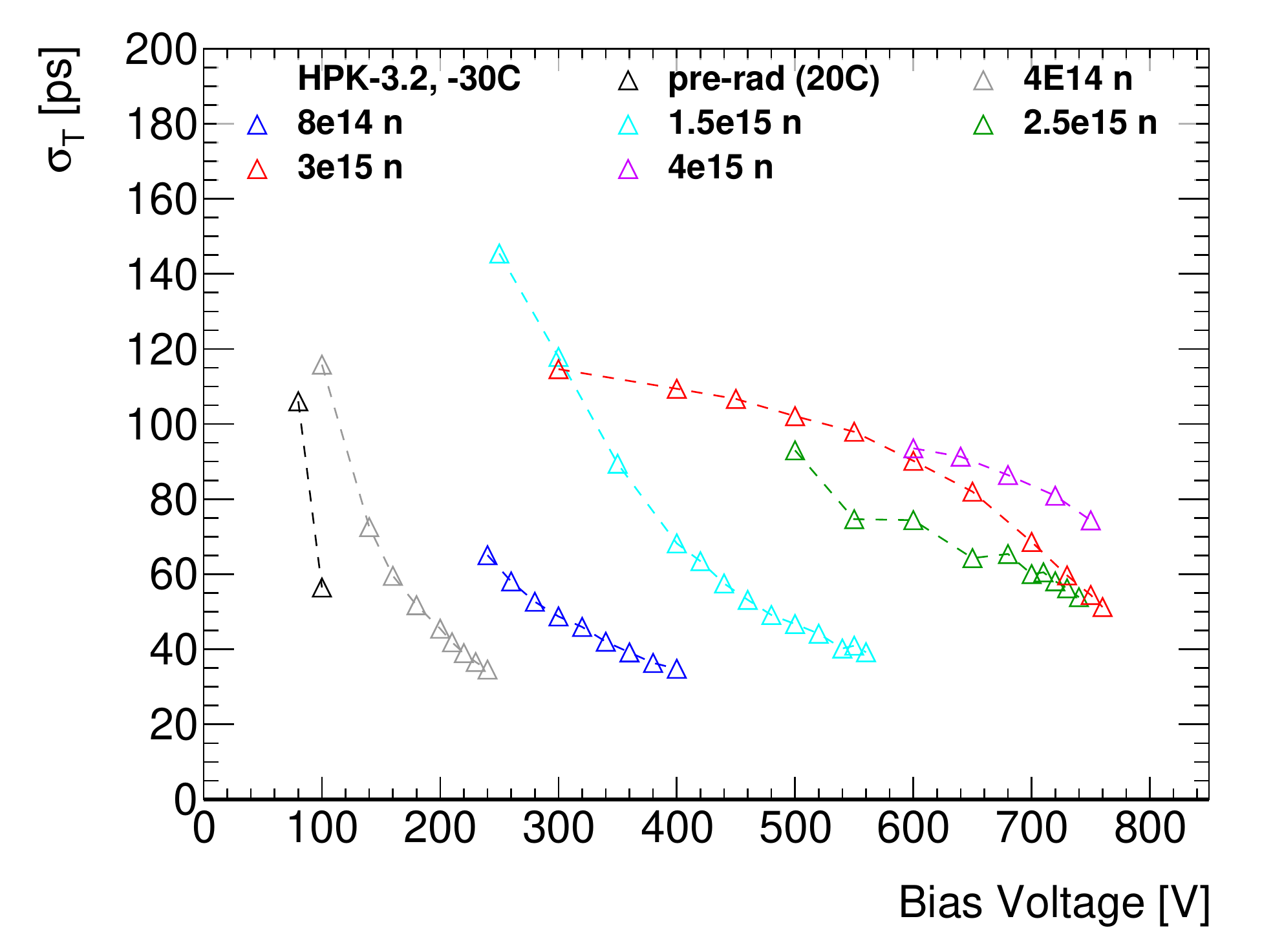}
\includegraphics[width=0.49\textwidth]{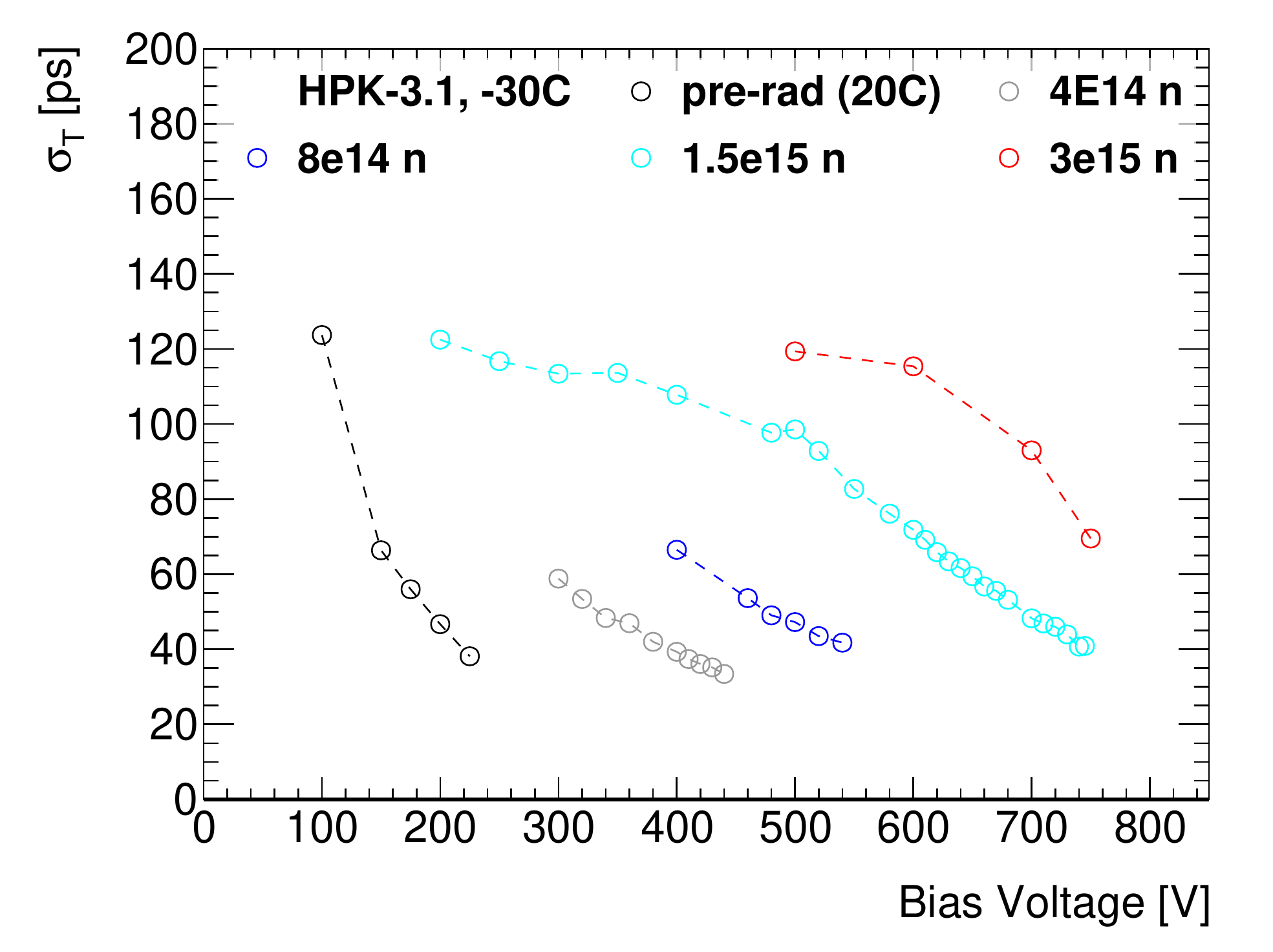}\\
\includegraphics[width=0.49\textwidth]{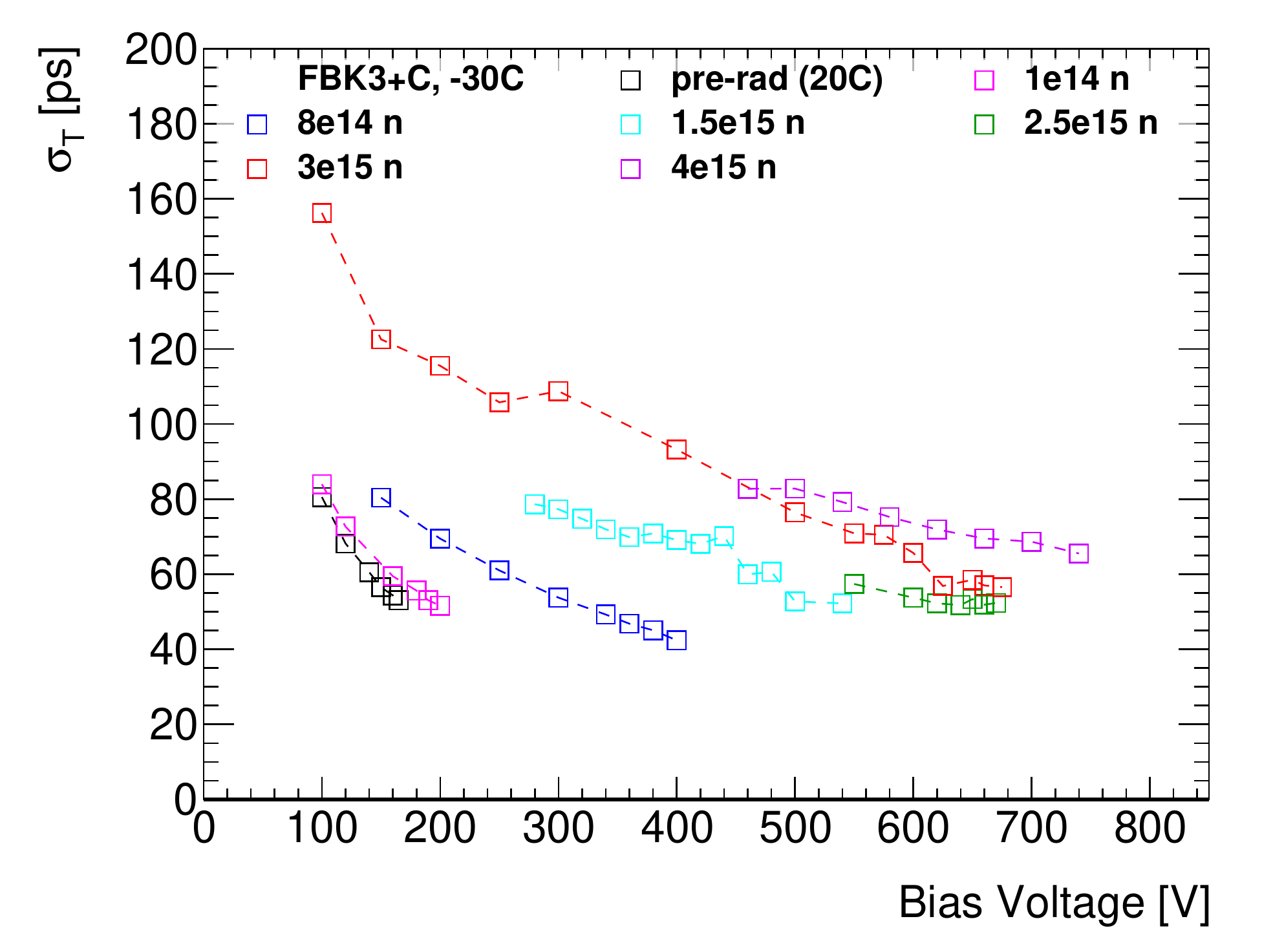}
\includegraphics[width=0.49\textwidth]{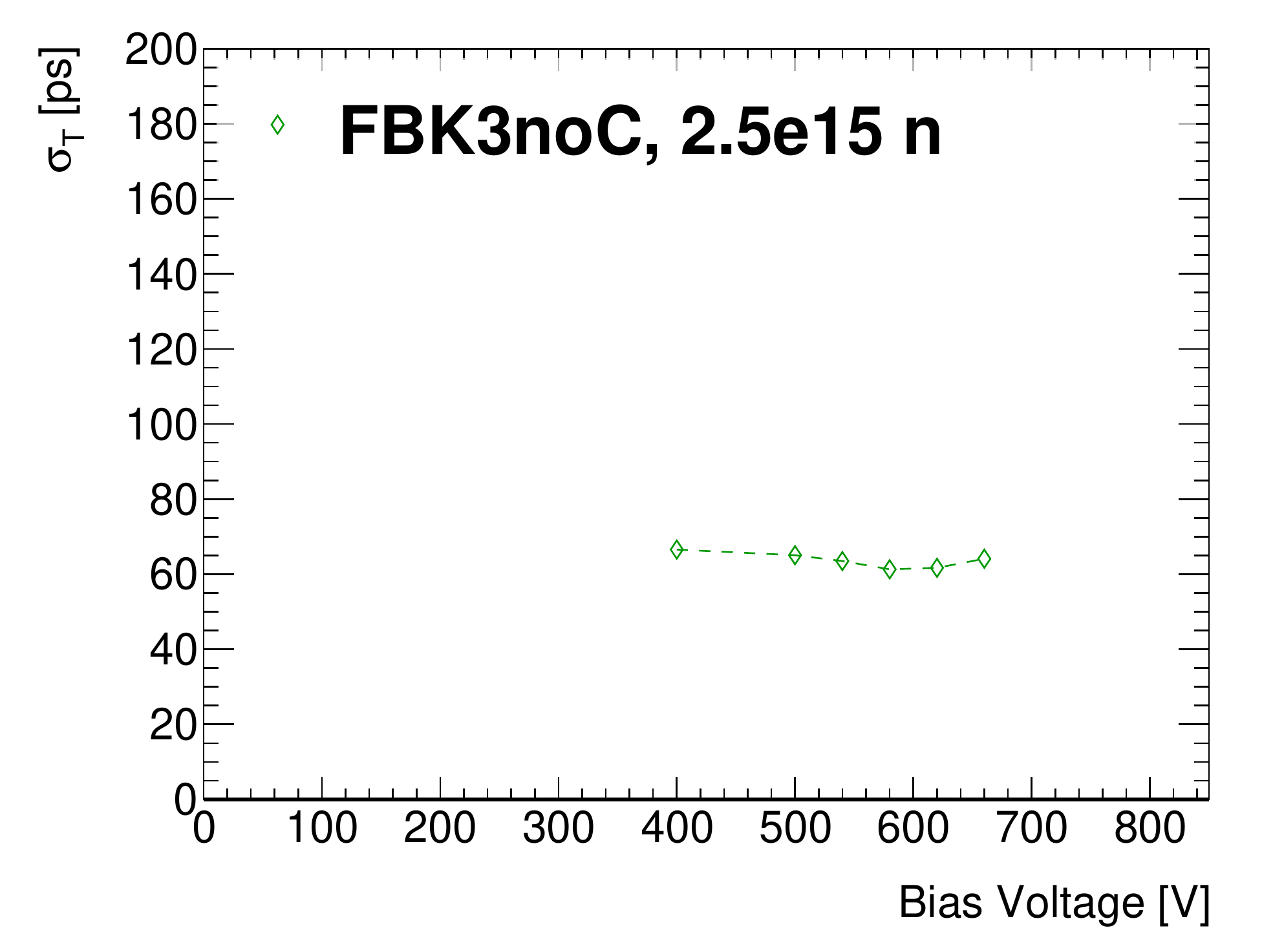}
\caption{Time resolution as a function of bias voltage for HPK-3.2 (top-left), HPK-3.1 (top-right), FBK3+C (bottom-left) and FBK3noC (bottom-right) sensors.}
\label{fig:timeres_beta}
\end{figure}

The time resolution is shown in \cref{fig:timeres_beta} for the four types of sensors.
For HPK-3.2, the time resolution before irradiation is fairly high for a \SI{50}{\micro\metre} thick LGAD; this effect is due to the excessive gain layer doping.
The performance is worse, around \SI{100}{\pico\second}, when the sensor is operated at cold temperatures because of the lowered breakdown voltage (around 75~V), which is very close to the full depletion voltage (70~V).
This behavior is consequence of the high initial doping of the multiplication layer, which leads to a breakdown voltage below the minimum voltage needed to obtain a saturated e/h drift velocity; for this reason the doping of future HPK-3.2 productions will be tuned to provide a good balance between operation before irradiation and radiation hardness.
The performance improves rapidly with acceptor removal, after slight radiation damage, at the fluence of \fluence{4}{14} the sensor can reach a time resolution of \SI{30}{\pico\second}.
At the fluence of \fluence{3}{15} HPK-3.2 can still reach a time resolution of less than \SI{60}{\pico\second}.
At higher fluences the time resolution is significantly worse.

HPK-3.1 has a better performance than HPK-3.2 before irradiation thanks to the higher operating voltage (breakdown voltage of 250~V and full depletion at 50~V). 
It has comparable time resolution to HPK-3.2 for low fluences, and can reach a time resolution around \SI{40}{\pico\second} up to a fluence of  \fluence{1.5}{15}.
For a fluence of \fluence{3}{15}, however, the time resolution is above \SI{60}{\pico\second}.

FBK3+C sensors has a slightly higher time resolution than HPK-3.1 and HPK-3.2, about \SI{40}{\pico\second} at lower fluences. It can reach a time resolution below than \SI{60}{\pico\second} up to \fluence{3}{15}. 
However, similarly to the case of HPK-3.2, at \fluence{4}{15} the time resolution increases above \SI{60}{\pico\second}. 

FBK3noC at \fluence{2.5}{15} has a time resolution of around \SI{60}{\pico\second}.

\FloatBarrier

\begin{figure}[htbp]
\centering
\includegraphics[width=0.7\textwidth]{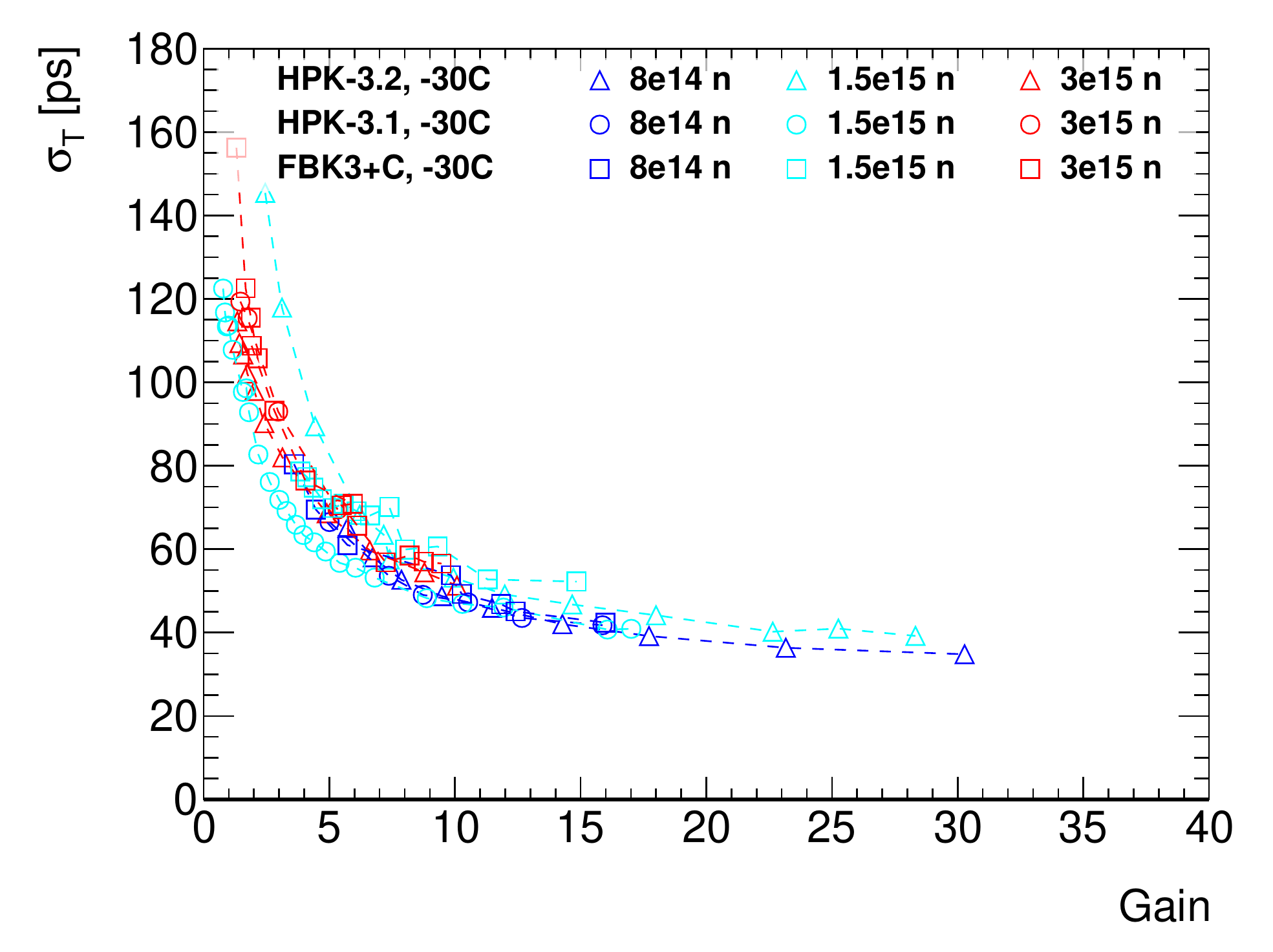}
\caption{Time resolution as a function of gain for HPK-3.2, HPK-3.1 and FBK3+C at different fluences.}
\label{fig:timeres_CC_beta}
\end{figure}

In \cref{fig:timeres_CC_beta} the time resolution is plotted as a function of gain for three types of sensors (HPK-3.2, HPK-3.1 and FBK3+C) each at three fluences (\fluence{8}{14}, \fluence{1.5}{15} and \fluence{3}{15}). 
It can be seen that all curves approximately lie on top of each other, showing a universal dependence of the time resolution with the gain across the studied sensor types, vendors and fluences.

\FloatBarrier
\section{CV and comparison with collected charge}
\label{sec:CV_comparison_CC}

\begin{figure}[htbp]
\centering
\includegraphics[width=0.49\textwidth]{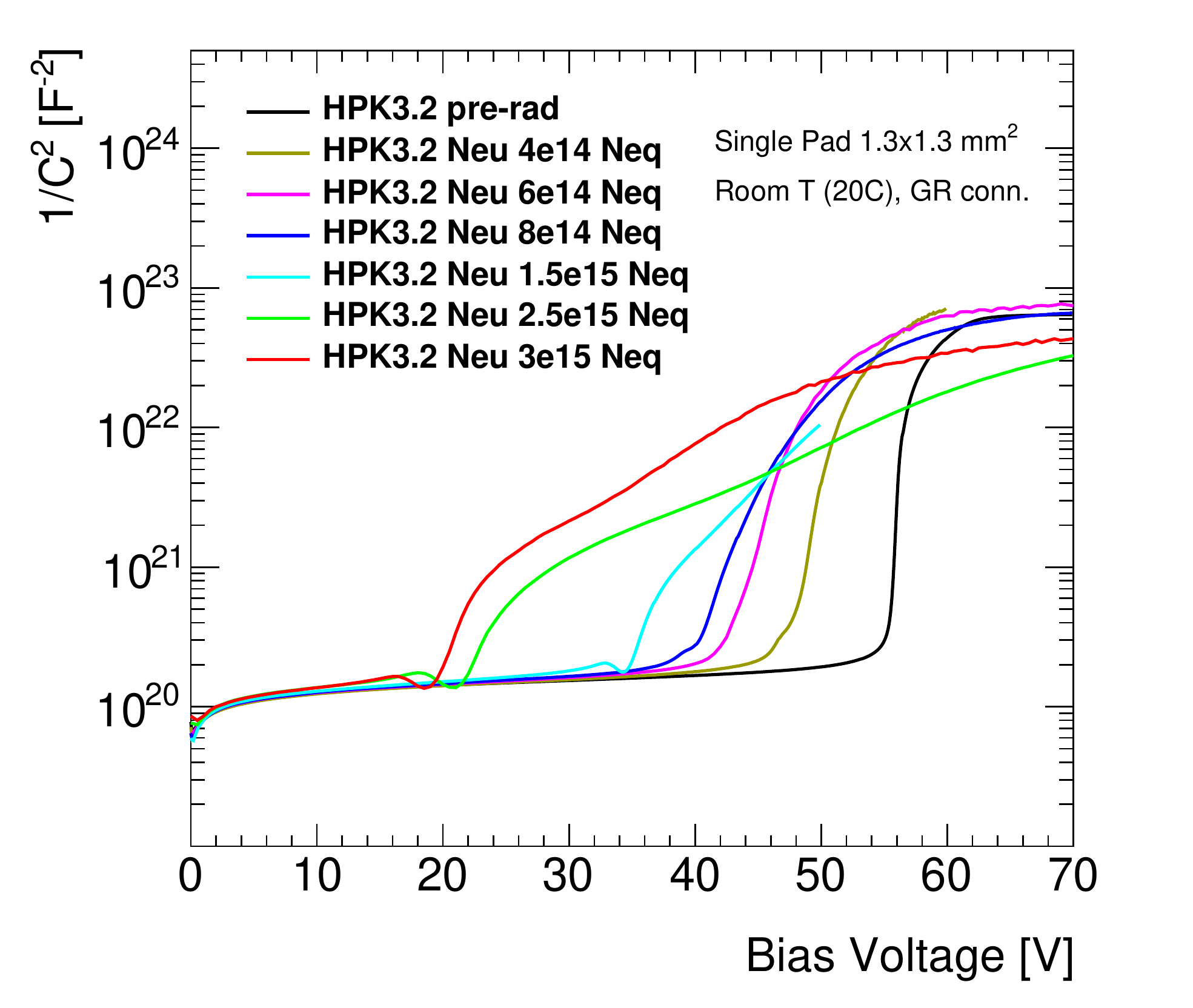}
\includegraphics[width=0.49\textwidth]{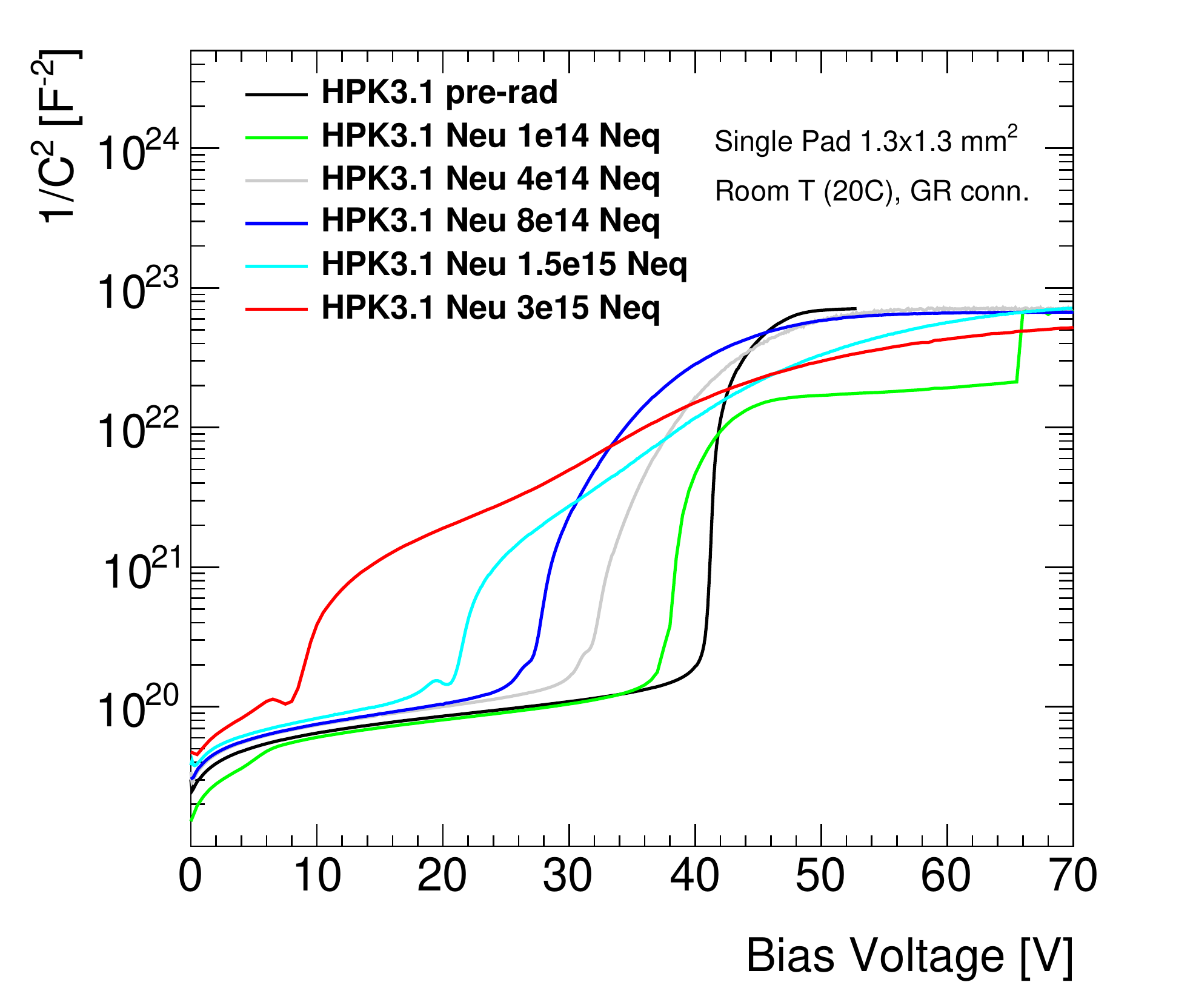}\\
\includegraphics[width=0.49\textwidth]{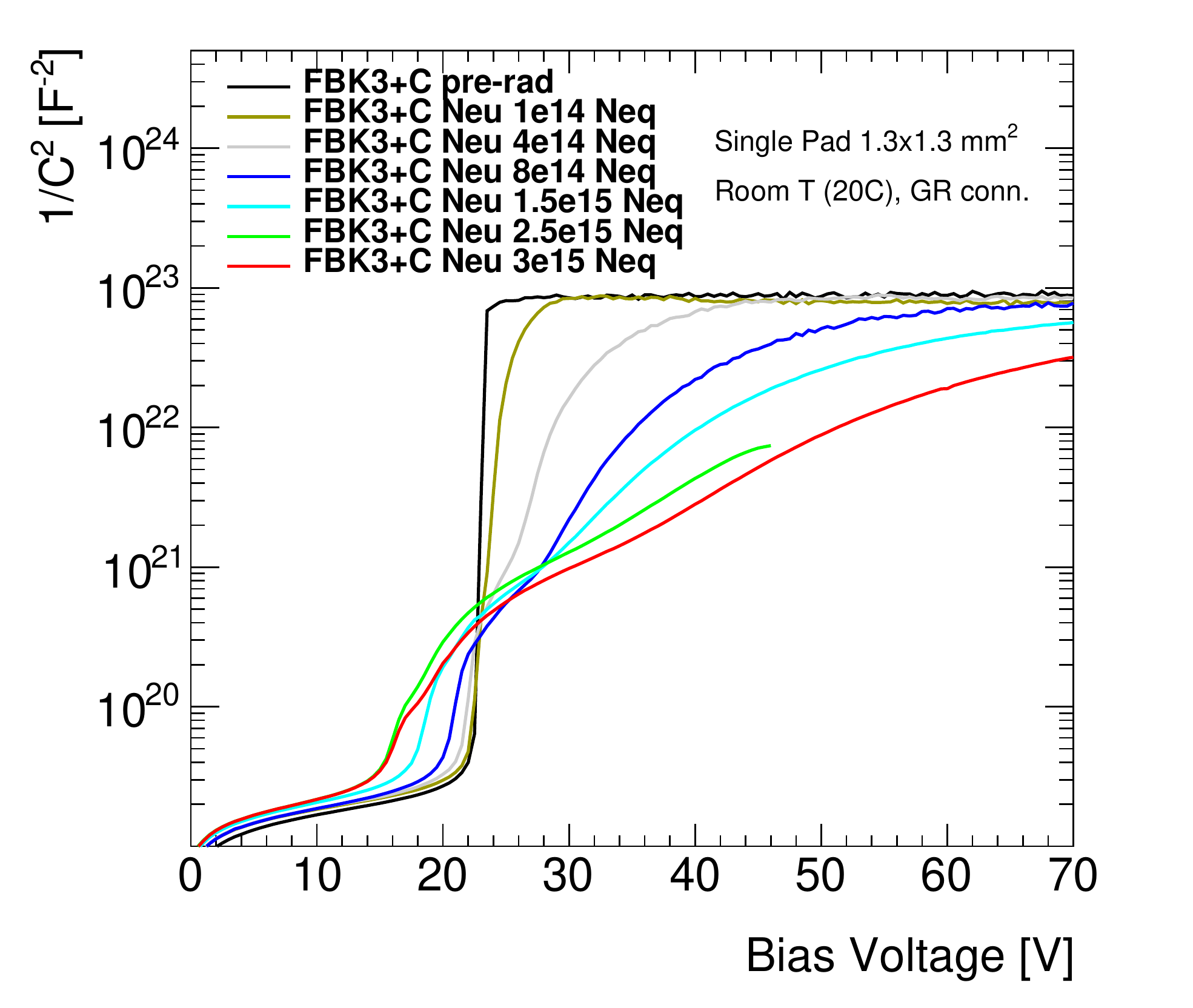}
\caption{C-V of HPK-3.2 (top-left), HPK-3.1 (top-right) and FBK+C (bottom) sensors at different radiation level.}
\label{fig:CV_fluence}
\end{figure}

As mentioned earlier, the doping concentration of the gain layer is proportional to the $V_{GL}$ extracted from the C-V measurement.
The $1/C^2$ measurement for HPK-3.2, HPK-3.1 and FBK3+C before and after irradiation can be seen in \cref{fig:CV_fluence}.
The effects of acceptor removal in the multiplication layer and the acceptor creation in the bulk are clearly visible by the shortening of $V_{GL}$ and the changes in the subsequent slope.

\begin{figure}[htbp]
\centering
\includegraphics[width=0.9\textwidth]{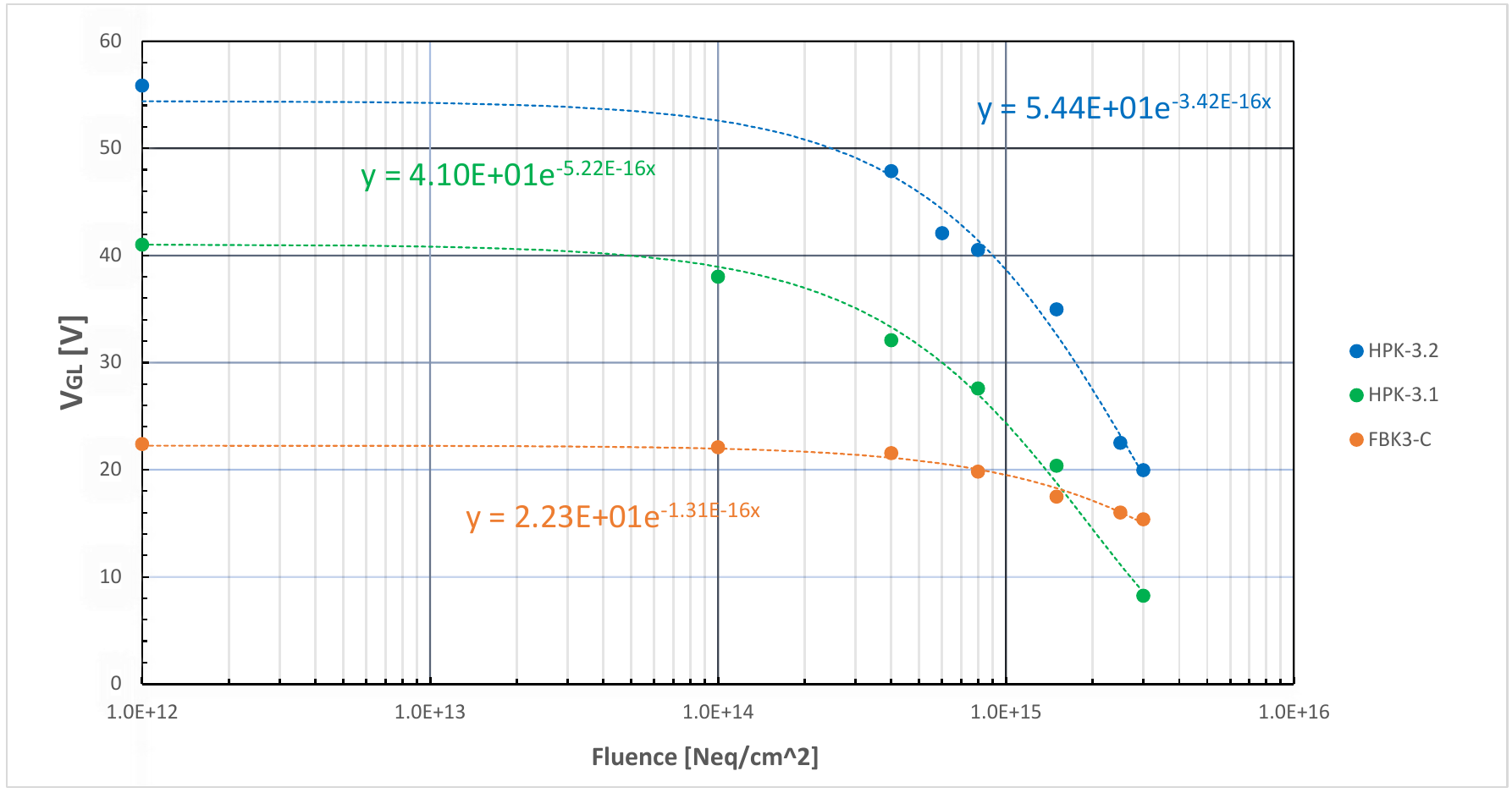}
\caption{Evolution of $V_{GL}$ (``foot'') for HPK-3.2, HPK-3.1 and FBK3+C sensors. Distribution for each sensor type are fitted with $N_D=N_0\cdot e^{-c\phi}$ to get the c coefficient.}
\label{fig:fraction_gain}
\end{figure}

From the $V_{GL}$ measurement, the amount of gain layer still active after a given fluence can be extracted.
In \cref{fig:fraction_gain}, $V_{GL}$ is showed as a function of the fluence $\phi$. 
The reduction of $V_{GL}$ can be fitted with the following formula to calculate the $c$ factor that represent the speed of doping removal:

\begin{equation}
N_D=N_0\cdot e^{-c\phi}
\end{equation}

Large differences are seen among the different sensor types: The HPK-3.1 has the largest c-factor ($c=5.22\cdot10^{16}$), about a factor 2 larger than HPK-3.2 ($c=3.42\cdot10^{16}$).
This shows the advantage of the deeper doping profile.
The FBK3+C has the smallest c coefficient ($c=1.31\cdot10^{16}$), more than a factor 2 smaller than HPK-3.2: we attribute this effect to the suppression of acceptor removal through the addition of carbon.

\begin{figure}[htbp]
\centering
\includegraphics[width=0.9\textwidth]{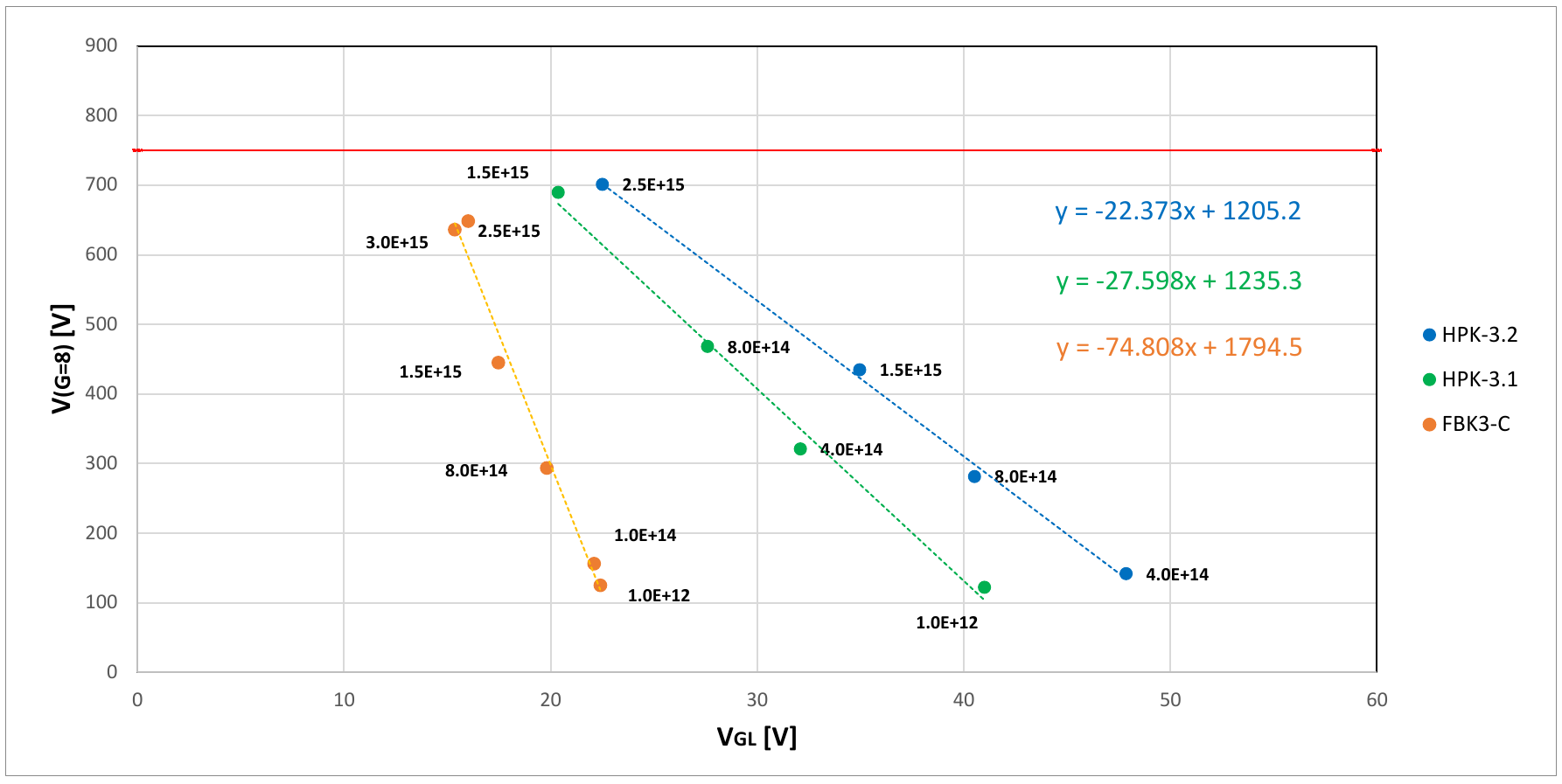}
\caption{
Correlation between $V_{GL}$ and the bias voltage to acquire gain of 8 V(G=8). 
Indicative linear fits are also shown in the plot.
Results are for three sensors from pre-rad (represented by \fluence{1}{12}) up to \fluence{3}{15}.
Only sensors that can reach a gain of 8 before 750~V are taken into consideration.
The red horizontal line represents 750~V which is the maximum safe voltage for \SI{50}{\micro\metre} thick sensors.}
\label{fig:gain_CV}
\end{figure}

The charge collection data (\cref{fig:CC_beta} and \cref{fig:CC_compare_beta}) can be characterized by the bias voltage needed to reach a certain gain G, V(G).
In the following analysis, the value to reach a gain of 8 (V(G=8), shown by the red line in \cref{fig:CC_beta} and \cref{fig:CC_compare_beta}) is taken into consideration.
This value can be correlated with the $V_{GL}$ value (in \cref{fig:fraction_gain}) measured in $1/C^2$ distributions shown in \cref{fig:CV_fluence}.
In the plot in \cref{fig:gain_CV} only sensors that can reach a gain of 8 are taken into consideration.
This correlation, seen in \cref{fig:gain_CV}, is fitted with an indicative linear function.
Both $V_{GL}$ and V(G=8) degrade altogether with radiation damage and show a fairly accurate linear correlation.
This correlation also means that, after a calibration of the method, the bias required to reach a gain after a certain fluence for a specific type of sensor can be derived from the calculation of $V_{GL}$ from a $1/C^2$ measurement.

\FloatBarrier
\section{Review of fluence uncertainty at JSI}

\begin{figure}[htbp]
\centering
\includegraphics[width=0.49\textwidth]{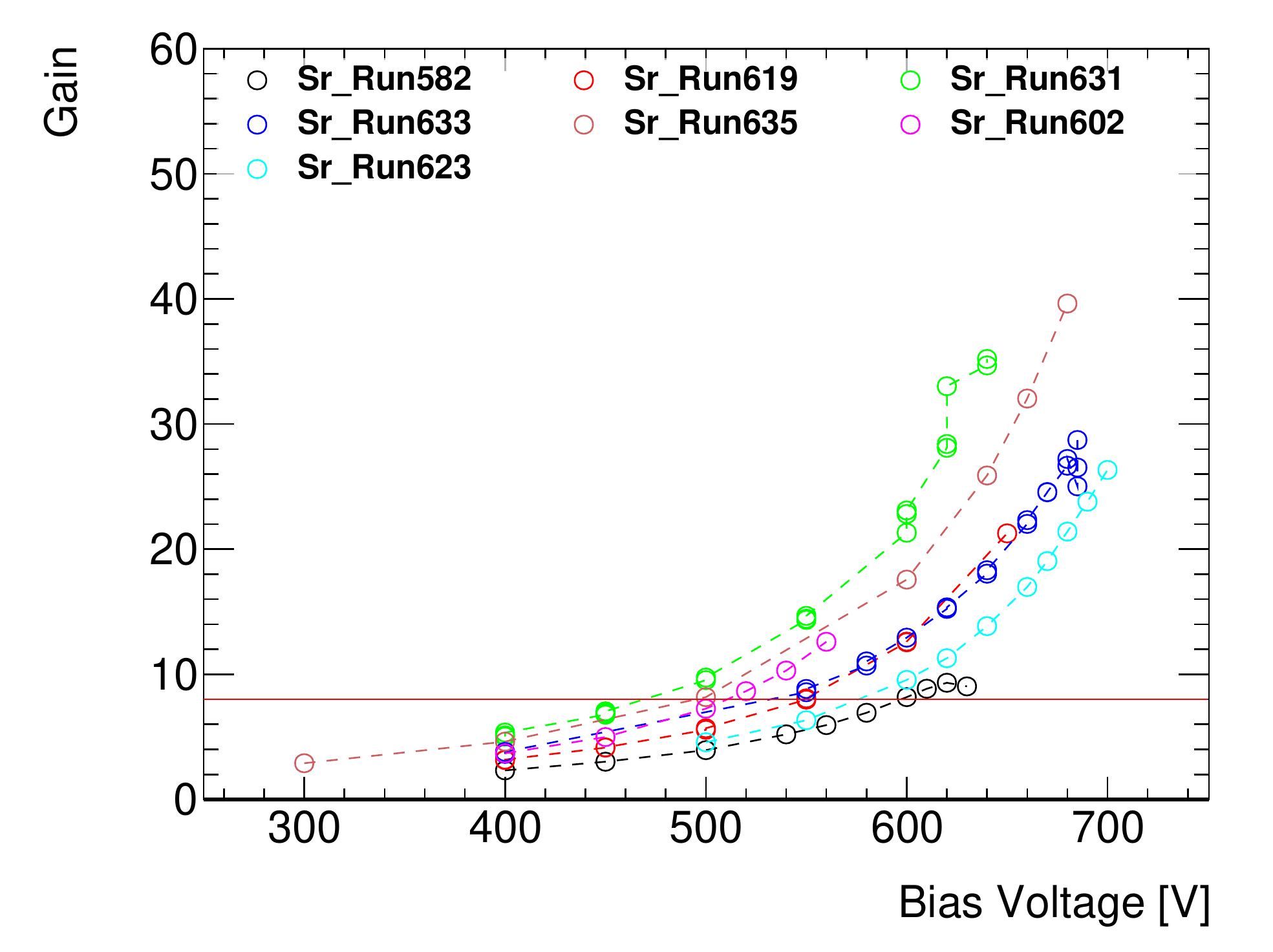}
\includegraphics[width=0.49\textwidth]{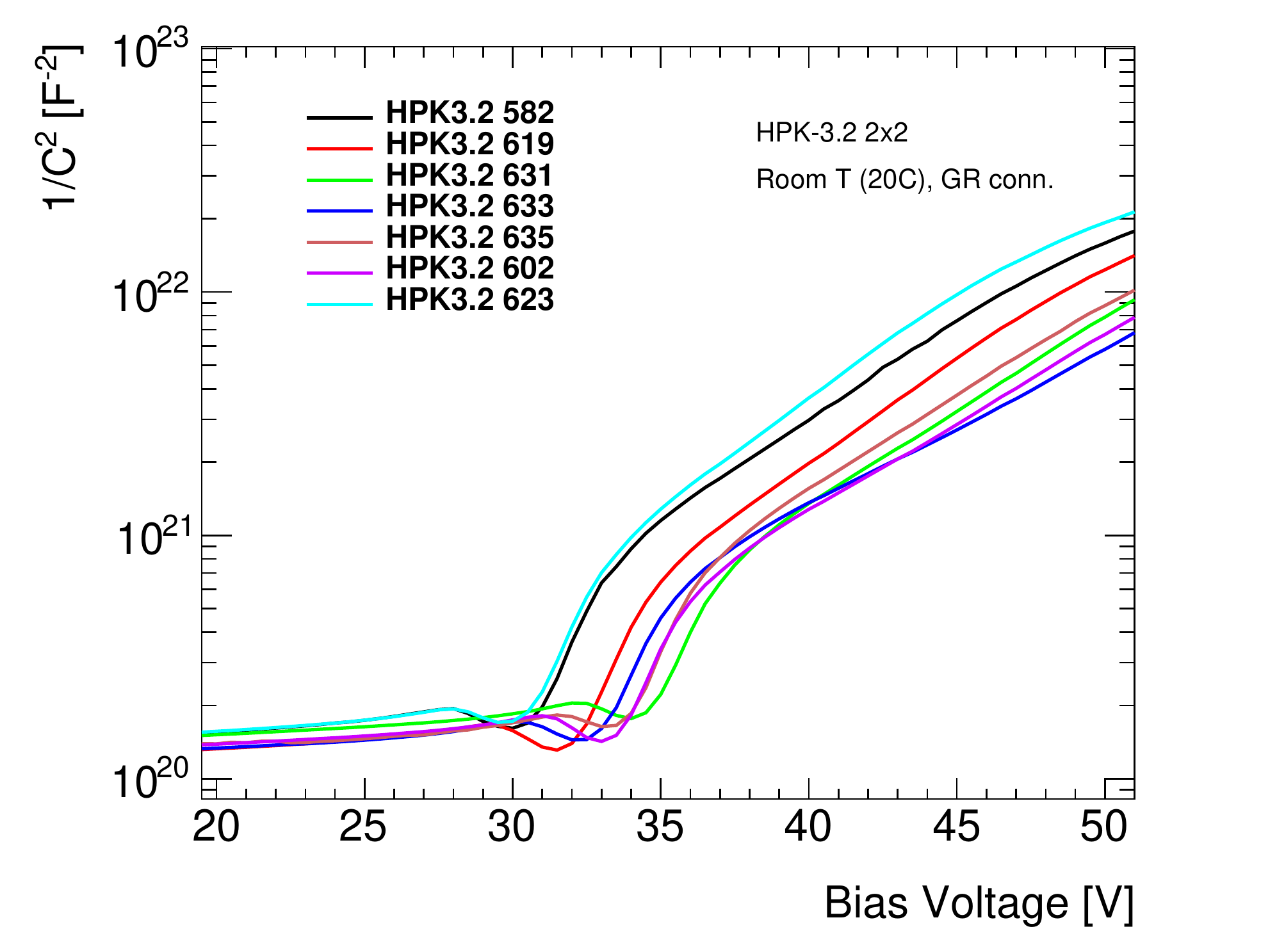}
\caption{Left: Difference in charge collection curves for several HPK-3.2 2x2 sensors all irradiated at a nominal fluence of \fluence{1.5}{15}. Different Run number (ID of the charge collection data) represent different HPK-3.2 sensors.
All sensors were irradiate together, however the positioning of each sensor was different inside the irradiation shuttle.
It can be seen that V(G=8) has roughly a 10\% variation.
Right: 1/$C^2$ distributions of the same HPK-3.2 sensors, also showing a variation in the $V_{GL}$ (``foot'') value.}
\label{fig:cc_spread}
\end{figure}

\begin{figure}[htbp]
\centering
\includegraphics[width=0.9\textwidth]{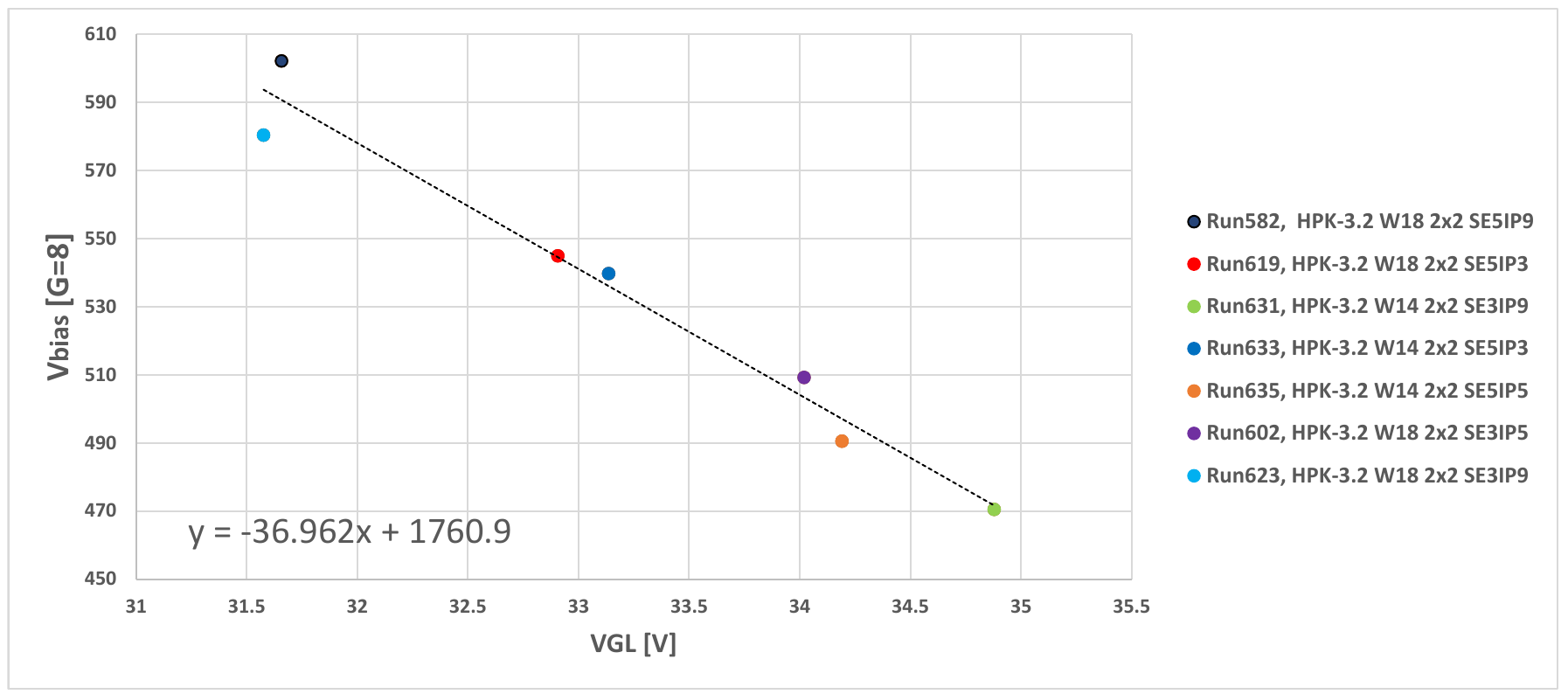}
\caption{Correlation plot between $V_{bias}$ to achieve gain 8, V(G=8), and the ``foot'', $V_{GL}$. An indicative linear fit is also shown in the plot.}
\label{fig:cc_spread_correlation}
\end{figure}

Several sensors for each type (HPK-3.1, HPK-3.2 and FBK3+C) were irradiated to the same fluence at JSI Ljubljana.
After sensors testing, a large spread of performance among identical sensors
irradiated to what was nominally the same fluence was found. 
As an example, \cref{fig:cc_spread} (left) shows the gain–bias curves for 7 different HPK-3.2 2x2 arrays of \SI[product-units=power]{1.3x1.3}{\mm} pads after a nominal neutron fluence of \fluence{1.5}{15}. The CV and collected charge measurements were made consistently on one of the pads of the 2x2 array.
In the plots the Run number (ID of the charge collection data) represents different individual sensor.
In \cref{fig:cc_spread} (right) is the 1/$C^2$ curve for the same sensors.
Every sensor was packed individually and put in the irradiation facility at the same time, but the position of each sensor was different inside the irradiation shuttle.
The variation in performance is evident: choosing the value V(G=8) as representative of the variation the spread in voltage is found to be of the order of 10\%.
The correlation of the V(G=8) value with $V_{GL}$ from $1/C^2$ (\cref{fig:cc_spread_correlation}) show a very linear behavior: this is a confirmation that the gain layer of the sensors tested is actually different and performance spread across the sensors is real, caused by different degrees of acceptor removal.

The data shown in \cref{fig:cc_spread_correlation} are for the same type of sensor (HPK-3.2) in \cref{fig:gain_CV}, however the geometry is different.
\cref{fig:gain_CV} contains data for single pad sensors while in \cref{fig:cc_spread_correlation} data for 2x2 arrays are shown, in both cases the pad tested has active area of \SI[product-units=power]{1.3x1.3}{\mm}.
The reason of the discrepancy between the fitted slopes in the two cases is unclear but can be related to the large fuence span in \cref{fig:gain_CV}, which influences the fit, and the different in geometry. 
Further investigation on this discrepancy will be done in the future.

\begin{figure}[htbp]
\centering
\includegraphics[width=0.6\textwidth]{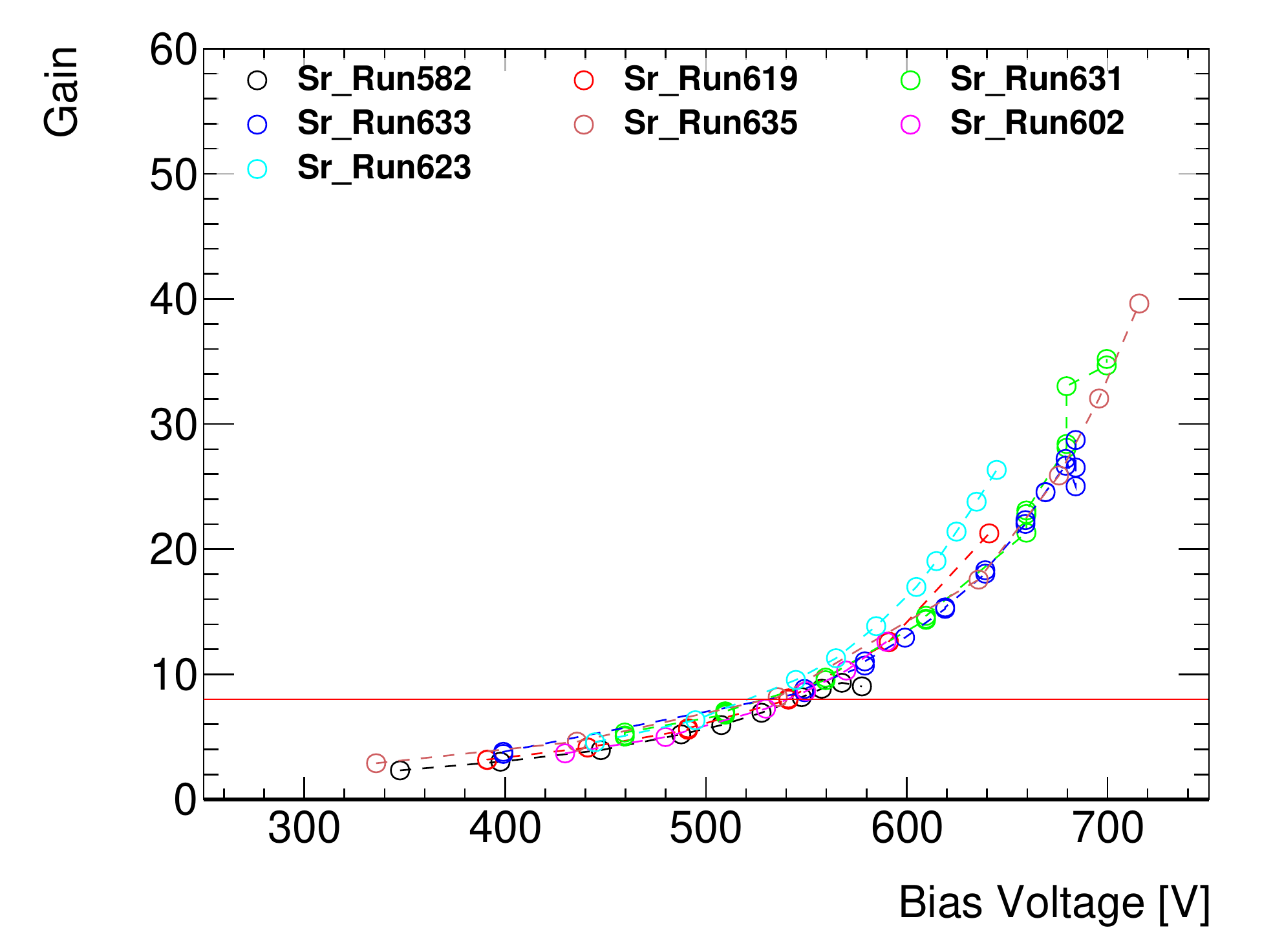}
\caption{Difference in charge collection curves for several HPK-3.2 sensors all irradiated at a nominal fluence of \fluence{1.5}{15} after applying a correction based on the $V_{GL}$ in \cref{fig:cc_spread_correlation}. Different Run number (ID of the charge collection data) represent different HPK-3.2 sensors. V(G=8) variation is around 1\% as was the case for sensors of the same type before irradiation.}
\label{fig:cc_spread_corr}
\end{figure}

A correction can be applied to the charge collection curves by using the $V_{GL}$ results from $1/C^2$ in the following manner: 
as first step the median value of $V_{GL}^M$ is calculated, this value is considered to be close to the sensor performance at the actual nominal fluence.
Afterwards the shift between the $V_{GL}$s and the median $V_{GL}^M$ is evaluated and correction factors are calculated for the $V_{bias}$ in the charge collection distributions form the linear correlation in \cref{fig:cc_spread_correlation}.
After this correction, the performance of the sensors match within a few percent (\cref{fig:cc_spread_corr}) equivalently to the spread these sensors had before irradiation.

The 10\% spread calculated seen in \cref{fig:cc_spread} is actually very close to the uncertainty value quoted by the JSI irradiation facility on the fluence.
The conclusion of this study is the confirmation that the sensor performance spread seen after irradiation is caused by the uncertainty on the fluence and not by reason intrinsic to the sensor type itself.

\FloatBarrier
\section{Conclusions}

The performance before and after irradiation of 2 types of HPK and 2 types of FBK sensors have been reviewed.
The sensors have been characterized electrically with I-V and C-V curves. 
From the C-V curves, an indication of the doping concentration and deepness of the gain layer is extracted using the ``foot'' or $V_{GL}$.
The study of charge collection has been done using the Sr-90 $\beta$-telescope setup at SCIPP, UCSC.
The following statements can be made from \cref{fig:CC_beta}, \cref{fig:CC_compare_beta} and \cref{fig:timeres_beta}:

\begin{itemize}
\item HPK-3.1 shows acceptable radiation hardness up to a neutron fluence of \fluence{1.5}{15}
    \item The very deep and thin gain layer of HPK-3.2 results in good radiation hardness up to a neutron fluence of \fluence{3}{15}, however the doping concentration needs to be tuned for optimal operation before irradiation.
    \item FBK sensors with Carbon (FBK3+C) implantation show exceptional radiation hardness compared to the same kind of sensors with no Carbon (FBK3noC) as seen in \cref{fig:CC_compare_beta} (right).
    \item HPK-3.2 and FBK3+C both show reasonable performance up to \fluence{3}{15}, however none of the presented types of sensors is radiation hard up to \fluence{4}{15}.
    \item Since the two technologies (Carbon in FBK3+C and deep gain layer in HPK-3.2) are independent from each other, a combination of a deep narrow gain layer with carbon can be expected to show significantly better  radiation hardness than either alone.
\end{itemize}

\noindent Further general conclusions:
\begin{itemize}
    \item The correlation between the time resolution and the gain shown in \cref{fig:timeres_CC_beta} is independent from the sensor type (with a similar thickness) and from the fluence.
    \item The depletion voltage of the gain layer ($V_{GL}$) and the voltage to reach a gain of 8 V(G=8) are linearly correlated.
    \item The correlation shows that it is possible, after careful calibration, to foresee the performance of LGADs after irradiation purely from a C-V scan.
    \item This technique can also be used to evaluate the fluence uncertainty of an irradiation facility.
\end{itemize}

\section{Acknowledgements}
This work was supported by the United States Department of Energy, grant DE-FG02-04ER41286, and partially performed within the CERN RD50 collaboration.
Part of this work has been financed by the European Union's Horizon 2020 Research and Innovation funding program, under Grant Agreement no. 654168 (AIDA-2020) and Grant Agreement no. 669529 (ERC UFSD669529), and by the Italian Ministero degli Affari Esteri and INFN Gruppo V.

\bibliography{bib/TechnicalProposal,bib/hpk_fbk_paper,bib/HGTD_TDR}

\clearpage
\appendix
\section{Direct gain comparison}
\label{sec:direct_comp}
\begin{figure}[htbp]
\centering
\includegraphics[width=0.45\textwidth]{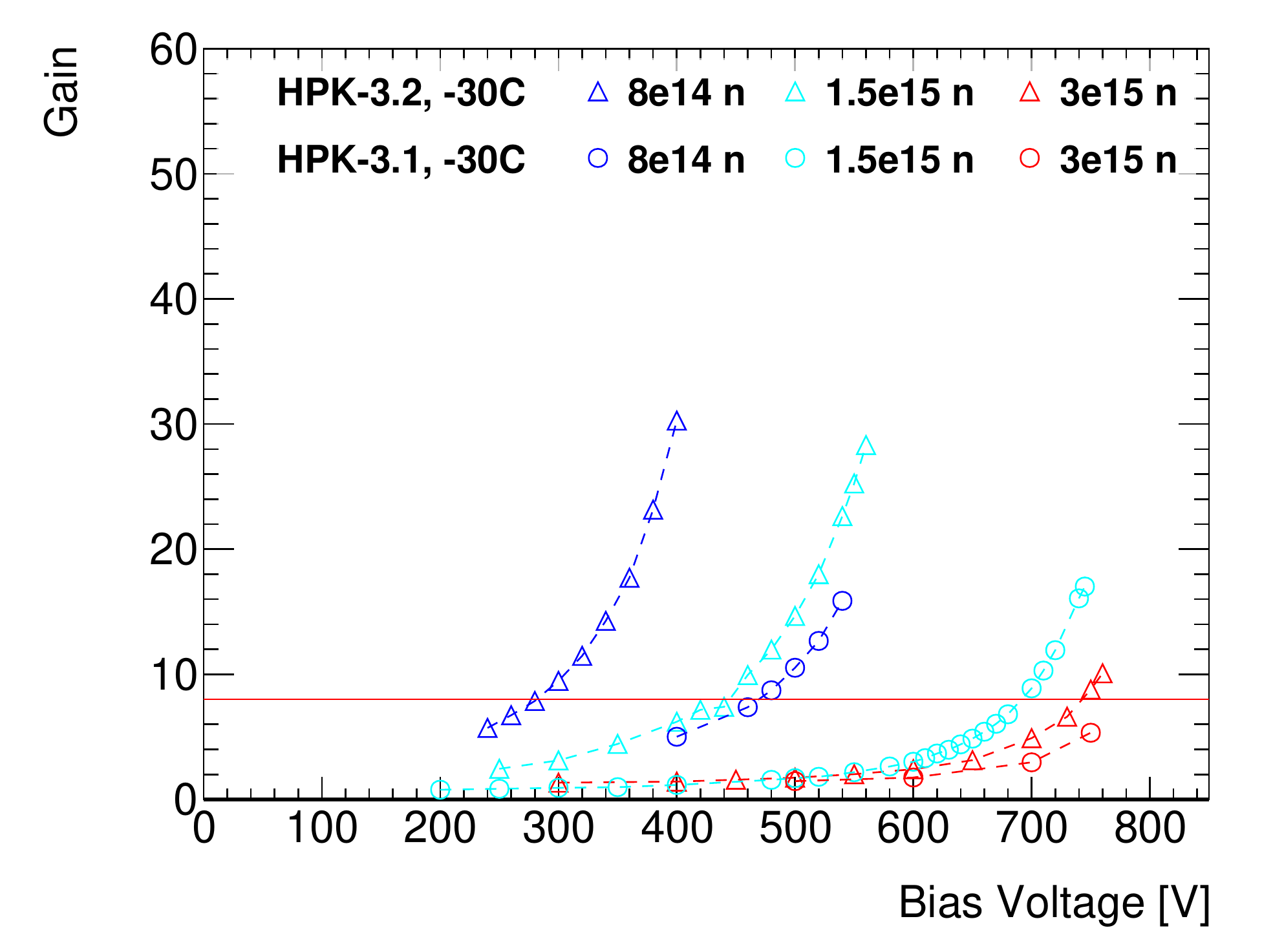}
\includegraphics[width=0.45\textwidth]{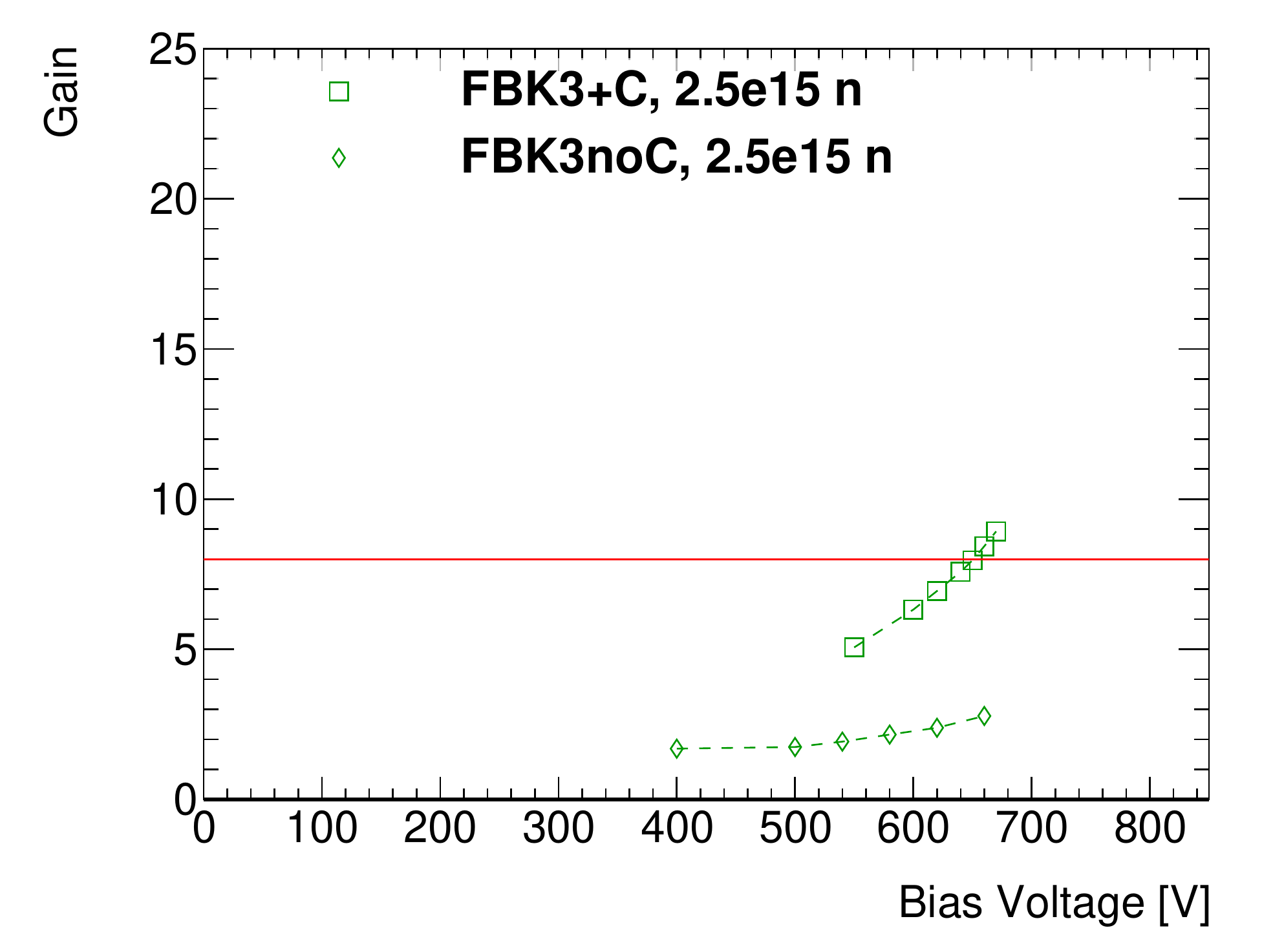}
\caption{Gain as a function of bias voltage for HPK-3.1/HPK-3.2 (left) and FBK3+C/FBK3noC (right) sensors. The two plots have different vertical scale.}
\label{fig:CC_compare_beta}
\end{figure}

\end{document}